\documentclass{aa}

\usepackage[varg]{txfonts}
\usepackage[colorlinks=true, linkcolor=blue, citecolor=blue, urlcolor=gray]{hyperref}
\usepackage{xcolor}
\usepackage{listings}

\bibpunct{(}{)}{;}{a}{}{,}

\defcitealias{2001ApJ...563..694V}{V01}
\defcitealias{2017MNRAS.469.2387P}{P17}
\defcitealias{2020ApJ...896....3K}{K20}

\begin{document}

\title{The multi-wavelength Tully-Fisher relation in the \\
TNG50 cosmological simulation}

\titlerunning{TNG50 Tully-Fisher relation}

\author{%
Maarten~Baes\inst{\ref{UGent}}
\and
Andrea~Gebek\inst{\ref{UGent}}
\and
Sabelo~Kunene\inst{\ref{UWC}}
\and
Lerothodi~Leeuw\inst{\ref{UWC}}
\and
Dylan~Nelson\inst{\ref{Heidelberg}}
\and
\\
Anastasia~A.~Ponomareva\inst{\ref{Hertfordshire},\ref{Oxford}}
\and
Nick~Andreadis\inst{\ref{UGent}}
\and
Alessandro~Bianchetti\inst{\ref{Padova},\ref{INAF-Padova}}
\and
W.~J.~G.~de~Blok\inst{\ref{ASTRON},\ref{UCT},\ref{Groningen}}
\and
\\
Sambatriniaina~H.~A.~Rajohnson\inst{\ref{UCT}}
\and
Amidou~Sorgho\inst{\ref{IAA}}
}

\institute{%
Sterrenkundig Observatorium, Universiteit Gent, Krijgslaan 281 S9, B-9000 Gent, Belgium\\ \email{maarten.baes@ugent.be}
\label{UGent}
\and
Department of Physics and Astronomy, University of the Western Cape, Robert Sobukwe Road, Bellville 7535, South Africa
\label{UWC}
\and
Institute of Theoretical Astrophysics, Center for Astronomy Heidelberg, Albert-\"Uberle-Stra{\ss}e 2, 69120 Heidelberg, Germany
\label{Heidelberg}
\and
Centre for Astrophysics Research, School of Physics, Astronomy and Mathematics, University of Hertfordshire, College Lane, Hatfield AL10 9AB, UK
\label{Hertfordshire}
\and
Oxford Astrophysics, Department of Physics, University of Oxford, Keble Road, Oxford OX1 3RH, UK
\label{Oxford}
\and
Dipartimento di Fisica e Astronomia, Universit\`a di Padova, Vicolo dell'Osservatorio 3, I-35122 Padova, Italy
\label{Padova}
\and
INAF -- Osservatorio Astronomico di Padova, Vicolo dell'Osservatorio 5, I-35122, Padova, Italy
\label{INAF-Padova}
\and
Netherlands Institute for Radio Astronomy (ASTRON), Oude Hoogeveensedijk 4, 7991 PD Dwingeloo, The Netherlands
\label{ASTRON}
\and
Department of Astronomy, University of Cape Town, Private Bag X3, Rondebosch 7701, South Africa
\label{UCT}
\and
Kapteyn Astronomical Institute, University of Groningen, PO Box 800, 9700 AV Groningen, The Netherlands
\label{Groningen}
\and
Instituto de Astrof\'{\i}sica de Andaluc\'{\i}a, Glorieta de la Astronom\'{\i}a s/n, 18008 Granada, Spain
\label{IAA}
}

\authorrunning{M.~Baes et al.}

\date{\today}

\abstract
{The Tully-Fisher relation (TFR) is one of the most important and widely used empirical correlations in extragalactic astronomy. Apart from its importance as a secondary distance indicator, the TFR relation serves as a test for galaxy evolution models, because it connects the baryonic and dark matter components of galaxies.}
{We aim at simulating the multi-wavelength TFR relation from UV to mid-infrared wavelengths for the TNG50 cosmological simulation at $z = 0$, and at comparing the results with observational TFR studies. We want to compare the wavelength dependence of the slope and scatter with the observed values, and search for secondary parameters that reduce the scatter in the TFR.}
{We select a large sample of simulated late-type, disc-dominated galaxies from the TNG50 simulation. For each galaxy, we use the SKIRT radiative transfer code to generate realistic synthetic global fluxes in 12 UV to mid-infrared broadbands and synthetic integrated H{\sc{i}} line profiles. We use bivariate linear regression to determine the TFR in each band, and we search for a second TFR parameter by correlating the residuals with different physical parameters.}
{Our TNG50 TFR reproduces the characteristic behaviour of the observed TFR in many studies: the TFR becomes steeper and tighter as we move from UV/optical to infrared wavelengths. The slope changes from $-7.46 \pm 0.14~{\text{mag}}~{\text{dex}}^{-1}$ in the NUV band to $-9.66 \pm 0.09~{\text{mag}}~{\text{dex}}^{-1}$ in the IRAC [4.5] band. Quantitatively, our slopes are well within the spread of different observational results. The ${\textit{u}} - {\textit{r}}$ colour or the sSFR can significantly reduce the scatter in the UV and optical bands. Using ${\textit{u}} - {\textit{r}}$ colour as second parameter, the modified TFR has a roughly constant intrinsic tightness of over the entire UV to MIR range.}
{The combination of the TNG50 cosmological simulation and the SKIRT radiative transfer postprocessing is capable of broadly reproducing the multi-wavelength TFR. A better matched sample selection, the use of different characteristic velocity scales, and more advanced internal dust attenuation correction are steps towards a more stringent comparison of the simulated and observed multi-wavelength TFR.}


\maketitle

%%%%%%%%%%%%%%%%%%%%%%%%%%%%%%%%%%%%%%%%%%%%%%%%%%%%%%

\section{Introduction}

The Tully-Fisher relation \citep[TFR,][]{1977A&A....54..661T} is an empirical relation that correlates the absolute magnitude and the rotation velocity of spiral galaxies. The TFR has been used for almost 50 years to estimate distances to spiral galaxies and to measure the Hubble constant \citep[e.g.,][]{1983ApJ...275..430V, 1988ApJ...331..620K, 1988ApJ...330..579P, 1992ApJ...387...47P, 2012ApJ...758L..12S, 2020AJ....160...71S, 2022MNRAS.511.6160K}. It is used for many other other cosmological applications too, such as determining the bulk flow of galaxies \citep{2009MNRAS.392..743W, 2023MNRAS.524.1885W, 2023MNRAS.526.3051W}, reconstructing the local Universe gravitational field in 3D \citep{2023A&A...670L..15C, 2023A&A...678A.176D}, and measuring the growth rate of cosmic structure at low $z$ \citep{2011MNRAS.413.2906D, 2015MNRAS.450..317C, 2024MNRAS.531...84B}. For a recent overview of the Tully-Fisher relation, in particular of its use as a distance indicator, we refer to \citet{Said2024}.

Apart from its practical application, the TFR is also important as a testbed for galaxy formation and evolution models. Indeed, the TFR is a very tight correlation, and any galaxy evolution model should reproduce its characteristics. Using the simple assumption that the circular velocity scales with the halo mass, one can argue that spiral galaxies should `naturally' lie along the TFR (\citealt{1998MNRAS.295..319M}; \citealt{1998ApJ...507..601V}; see Appendix B of \citealt{2007ApJ...671..203C}). Galaxy stellar masses or luminosities are, however, not directly proportional to the mass of the halo they inhabit, or in other words, galaxy formation efficiency is a non-monotonic function of halo mass \citep{2013MNRAS.428.3121M, 2013ApJ...770...57B, 2019MNRAS.488.3143B, 2018ARA&A..56..435W, 2019MNRAS.486.5468L}. For galaxies to lie along the TFR, the ratio of the rotation velocity and the circular velocity of galaxies should also non-monotonically depend on halo mass \citep{2017MNRAS.464.4736F}. This makes the TFR a more challenging test for galaxy formation models than this simple argument seems to suggest. 

This particularly applies to cosmological hydrodynamical models of galaxy formation, currently one of the most popular methods to study galaxy evolution \citep{2015ARA&A..53...51S, 2020NatRP...2...42V, 2023ARA&A..61..473C}. Up to about a decade ago, it was hard to simulate populations of spiral galaxies that reproduce the slope, zero-point and scatter of the TFR within the $\Lambda$CDM framework. In particular, early hydrodynamical simulations produced spiral galaxies that were too massive and too compact, and had steeply declining rather than flat rotation curves \citep{2003ApJ...591..499A, 2010MNRAS.408..812S, 2012MNRAS.423.1726S}. This resulted in a failure to reproduce the observed TFR \citep[e.g.,][]{2000ApJ...538..477N}. 

The culprit of this failure was the inability of early feedback schemes to prevent the accumulation of low angular momentum baryons at the centre of dark matter haloes. The inclusion of more fine-tuned prescriptions for feedback by stellar winds, supernovae and supermassive black holes turned out to be key to reproducing populations of more realistic spiral galaxies \citep[e.g.][]{2007MNRAS.374.1479G, 2011MNRAS.415.1051B, 2011ApJ...742...76G, 2012MNRAS.420.2245M, 2014MNRAS.437.1750M}. As a result, state-of-the-art cosmological hydrodynamical simulations now generate TFRs that agree relatively well with observations \citep{2014MNRAS.444.1518V, 2017MNRAS.464.2419S, 2017MNRAS.464.4736F, 2020MNRAS.498.3687G, 2021MNRAS.507.3267G, 2021A&A...651A.109D, 2023MNRAS.520.3895G}.

One important aspect that needs to be stressed is that all of the studies cited above actually do not simulate the `original' TFR, but rather the stellar TFR (STFR) and/or the baryonic TFR (BTFR). For the STFR, the absolute magnitude (or luminosity) is replaced by the stellar mass; for the BTFR, it is replaced by the total baryonic mass, usually interpreted as the sum of the stellar mass and the cold ISM gas mass \citep{2000ApJ...533L..99M, 2001ApJ...550..212B}. While it has been argued that the BTFR is more fundamental than the original TFR \citep{2012AJ....143...40M}, the main argument for choosing the STFR or BTFR when testing galaxy formation models is that stellar and baryonic mass are easily determined for simulated galaxies. The drawback is that they are harder to accurately measure for observed galaxies \citep{2018MNRAS.474.4366P}. The determination of stellar masses requires multi-band photometry, and stellar masses are then typically determined using SED fitting. Typical uncertainties on global stellar masses derived from integrated multi-band photometry are $\sim$0.3~dex \citep{2012MNRAS.422.3285P, 2013ARA&A..51..393C, 2015MNRAS.452.3209R, 2020ApJ...904...33L, 2023ApJ...944..141P}. Baryonic mass determinations need, in addition to stellar masses, estimates for the atomic and molecular gas mass. These can be determined from H{\sc{i}} 21cm observations and CO observations, respectively, with the poorly constrained CO-to-H$_2$ conversion factor \citep[e.g.][]{2013ARA&A..51..207B, 2016A&A...588A..23A, 2020A&A...643A.141M} an additional source of uncertainty.

Rather than, or in addition to, the STFR or BTFR, the original TFR forms an interesting test for galaxy formation models, especially if we consider that the TFR has been observed in different bands. Early measurements were mainly done at optical wavelengths \citep{1977A&A....54..661T, 1987ApJ...313..629B, 1997AJ....113...53G}, and the TFR was subsequently extended to the near-infrared \citep[NIR:][]{2000ApJ...533..781R, 2001ApJ...563..694V, 2002A&A...396..431K, 2007A&A...465...71T, 2008AJ....135.1738M} and to the mid-infrared \citep[MIR:][]{2013ApJ...765...94S, 2013ApJ...771...88L, 2014ApJ...792..129N, 2020ApJ...896....3K, 2023MNRAS.519..102B}. While there is a significant spread in the slope and scatter measurements among the different observational studies, a general trend is that the TFR becomes steeper and tighter as we move to longer wavelengths. This is demonstrated most convincingly by studies that apply a consistent and uniform approach at different wavelengths \citep[e.g.,][]{2000ApJ...529..698S, 2001ApJ...563..694V, 2017MNRAS.469.2387P, 2020ApJ...896....3K}. Moreover, it has been shown that, at least in optical bands, the scatter of the TFR correlates with other galaxy characteristics, such as colour or morphological type \citep{1983ApJ...265....1A, 1985ApJ...289...81R, 1997AJ....113...53G, 2004ApJ...607..241R, 2006ApJ...653..861M, 2017MNRAS.469.2387P, 2018MNRAS.479.3373M, 2020ApJ...896....3K}. Reproducing these different aspects of the multi-wavelength TFR is an interesting test for any galaxy formation model. 

Simulating the multi-wavelength TFR for a cosmological hydrodynamical model requires the generation of mock fluxes in different bands. This requires the combination of stellar population models and realistic models for dust attenuation. The latter aspect is particularly tricky: on average, interstellar dust attenuates about a third of all the stellar emission in spiral galaxies \citep{2002MNRAS.335L..41P, 2016A&A...586A..13V, 2018A&A...620A.112B}. The effects of dust attenuation are often complex and even counter-intuitive \citep{1992ApJ...393..611W, 1994ApJ...432..114B, 2004ApJ...617.1022P, 2006A&A...456..941M, 2010MNRAS.403.2053G}. Calculating accurate fluxes in the complex 3D geometry of simulated galaxies from hydrodynamic simulations can only be done using 3D radiative transfer calculations \citep{2013ARA&A..51...63S}. Until about a decade ago, this was still a significant challenge, but thanks to improved methods and computing power, this is now possible. In fact, radiative-transfer-based mock broadband fluxes and/or images have been computed for most state-of-the-art cosmological hydrodynamical simulations \citep[e.g.,][]{2015MNRAS.447.2753T, 2017MNRAS.470..771T, 2018ApJS..234...20C, 2022MNRAS.512.2728C, 2021MNRAS.506.5703K, 2022MNRAS.516.3728T, 2022ApJ...929...94A, 2023ApJ...957....7F, 2023MNRAS.519.4920G, 2023ApJ...950....4J, 2024MNRAS.531.3839G, 2024A&A...683A.181B, 2024MNRAS.527.6506B}.

\begin{figure*}
\includegraphics[width=\textwidth]{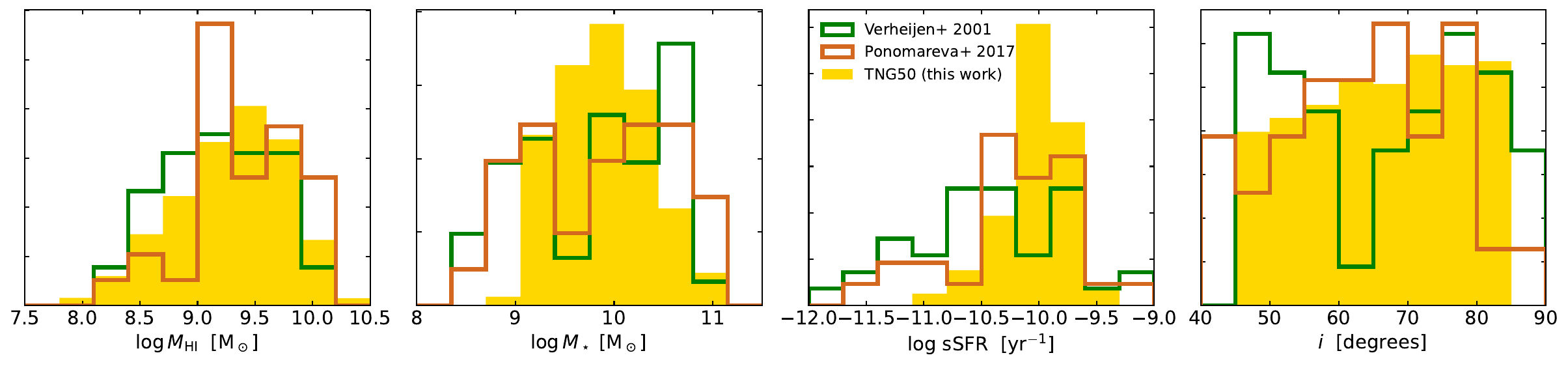}
\caption{Histograms of basic properties of our TNG50 and the \citetalias{2001ApJ...563..694V} and \citetalias{2017MNRAS.469.2387P} galaxy samples. From left to right: H{\sc{i}} mass, stellar mass, specific star formation rate, and inclination. For the simulated TNG50 galaxies, H{\sc{i}} masses and inclinations are determined as discussed in Sec.~{\ref{LineWidths.sec}}, while stellar masses and SFRs are obtained directly from the simulation particle data \citep{2019ComAC...6....2N}. For the \citetalias{2001ApJ...563..694V} sample, H{\sc{i}} masses and inclinations are obtained from \citet{2001A&A...370..765V}. Stellar masses are calculated using the relation between {\textit{K}}-band mass-to-light ratio and ${\textit{B}}-{\textit{R}}$ colour by \citet{2001ApJ...550..212B}, with ${\textit{B}}$, ${\textit{R}}$ and ${\textit{K}}$ absolute magnitudes taken from \citet{1996AJ....112.2471T}. SFRs for the \citetalias{2001ApJ...563..694V} sample galaxies are estimated from integrated {\it{GALEX}} FUV-band luminosities using Eq.~(3) of \citet{2009ApJ...706..599L}, based on fluxes collected from different sources, primarily \citet{2011ApJ...741..124H} and \citet{2018ApJS..234...18B}. For the \citetalias{2017MNRAS.469.2387P} sample, H{\sc{i}} masses and inclinations are directly taken from \citetalias{2017MNRAS.469.2387P}. For the galaxies of this sample, stellar masses and SFRs have been determined by \citet{2018MNRAS.474.4366P} using SED fits with the MAGPHYS code \citep{2008MNRAS.388.1595D}.}
\label{SampleHistograms.fig}
\end{figure*}

If we want to compare simulated to observed TFRs beyond the qualitative level, it is also important to pay attention to the measurement of the rotation velocity from the simulated galaxies. There are different possibilities to define a characteristic rotation speed for a simulated galaxy. \citet{2014MNRAS.444.1518V}, \citet{2017MNRAS.464.2419S}, and \citet{2017MNRAS.464.4736F} adopted the circular velocity,
\begin{equation}
v_{\text{c}}(r) = \sqrt{\frac{GM(r)}{r}},
\label{vc}
\end{equation}
evaluated at twice the stellar half-mass radius, at twice the baryonic half-mass radius, and the stellar half-mass radius, respectively. \citet{2020MNRAS.498.3687G, 2021MNRAS.507.3267G} considered four different metrics for the characteristic velocity based on the circular velocity curve (\ref{vc}). A more consistent approach is to mimic the observational approach by generating mock H{\sc{i}} data and determining the rotation velocity in the same way as done for observed galaxies. This approach was followed by \citet{2023MNRAS.520.3895G}, who used the MARTINI package \citep{2019MNRAS.482..821O} to create mock 21~cm spectra, allowing for a fair comparison between real and mock galaxies from the TNG100 simulation.

In this paper, we study the $z=0$ multi-wavelength TFR for the TNG50 cosmological hydrodynamical model \citep{2019MNRAS.490.3196P, 2019MNRAS.490.3234N}. This simulation combines {{the high resolution typical for zoom-in simulations}} with a volume that is large enough to host galaxy populations with sufficient statistics (for more details, see Sec.~{\ref{SampleSelection.sec}}). {{In this paper we derive the multi-wavelength TFR for a sample of simulated spiral galaxies from the TNG50 simulation at $z=0$. We pay special care for the calculation of the multi-wavelength luminosities and the rotation velocities of the simulated galaxies, mimicking the observational approach as closely as possible.}} Our goal is to compare the TNG50 multi-wavelength TFR {{to}} the one obtained from observational studies, particularly with studies that applied a uniform analysis in different bands, such as \citet[][hereafter \citetalias{2001ApJ...563..694V}]{2001ApJ...563..694V}, \citet[][hereafter \citetalias{2017MNRAS.469.2387P}]{2017MNRAS.469.2387P}, and \citet[][hereafter \citetalias{2020ApJ...896....3K}]{2020ApJ...896....3K}.
Very specifically, we want to investigate whether the TNG50 simulation can reproduce the slope and the tightness of the TFR obtained by these studies over the entire UV--MIR wavelength range, and whether a second parameter can be identified that can reduce the scatter in the TFR. 

This paper is organised as follows: in Sec.~{\ref{SampleSelection.sec}} we present our sample selection, in Sec.~{\ref{DataGeneration.sec}} we discuss how we generate the synthetic data projects needed, in Sec.~{\ref{TNG50-TF.sec}} we present the TNG50 multi-wavelength TFR, in Sec.~{\ref{Discussion.sec}} we discuss our results, and in Sec.~{\ref{Summary.sec}} we present our summary and an outlook.

%%%%%%%%%%%%%%%%%%%%%%%%%%%%%%%%%%%%%%%%%%%%%%%%%%%%%%

\section{The TNG50 sample selection}
\label{SampleSelection.sec}

The TNG50 simulation \citep{2019MNRAS.490.3196P, 2019MNRAS.490.3234N} is the smallest-volume and highest-resolution version of the IllustrisTNG suite of cosmological magneto-hydrodynamical simulations \citep{2018MNRAS.480.5113M, 2018MNRAS.477.1206N, 2018MNRAS.475..624N, 2018MNRAS.475..648P, 2018MNRAS.475..676S}. It is based on the moving-mesh hydrodynamics code AREPO \citep{2010MNRAS.401..791S} and a galaxy formation model that includes gas cooling and heating, stochastic star formation, stellar evolution, chemical enrichment of the ISM, feedback from supernovae, seeding and growth of supermassive black holes, AGN feedback, and magnetic fields \citep{2017MNRAS.465.3291W, 2018MNRAS.473.4077P}. The cosmological parameters ($\Omega_{\text{m}}= 0.3089$, $\Omega_{\text{b}}= 0.0486$, $\Omega_\Lambda = 0.6911$, and $H_0 = 67.74~{\text{km}}\,{\text{s}}^{-1}\,{\text{Mpc}}^{-1}$) are based on the Planck 2015 results \citep{2016A&A...594A..13P}. For a full description of the TNG50 simulation we refer to \citet{2019MNRAS.490.3196P} and \citet{2019MNRAS.490.3234N}.

\begin{figure*}
\includegraphics[width=\textwidth]{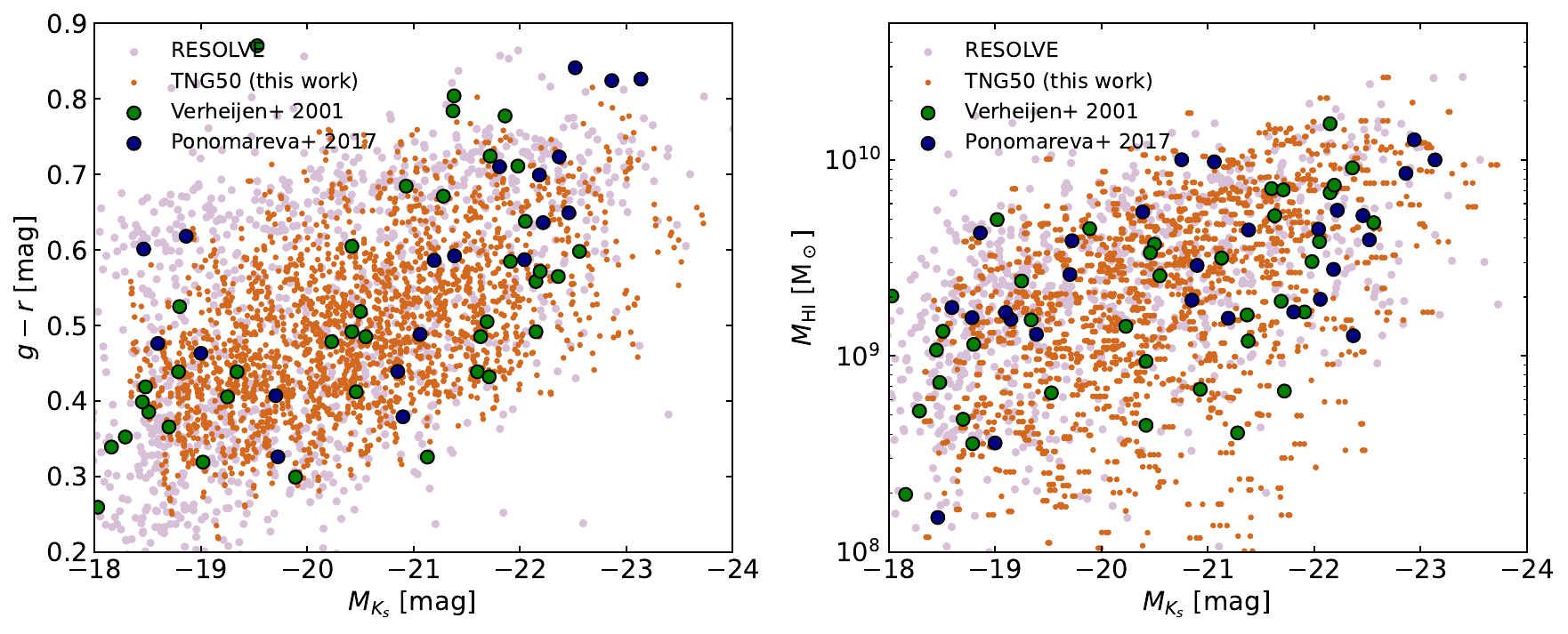}
\caption{Scatter plots comparing the location of our TNG50 simulated galaxy sample, the \citetalias{2017MNRAS.469.2387P} and \citetalias{2017MNRAS.469.2387P} galaxy samples, and the volume-limited RESOLVE galaxy sample. The left and right panel show the ${\textit{g}}-{\textit{r}}$ colour and the H{\sc{i}} mass as a function of ${\textit{K}}_{\text{s}}$-band absolute magnitude, respectively. Colours and absolute magnitudes for the simulated TNG50 galaxies are determined as described in Sect.~{\ref{AbsoluteMagnitudes.sec}}. For the \citetalias{2017MNRAS.469.2387P} sample, ${\textit{g}}-{\textit{r}}$ colours are calculated from the ${\textit{B}}-{\textit{R}}$ colours using the conversion formula of \citet{2011AJ....141...47F}.}
\label{RESOLVE.fig}
\end{figure*}

The aim of our study is to derive the multi-wavelength TFR of galaxies in the TNG50 simulation and compare it to the {{multi-wavelength TFR}} observed in the Local Universe. Our base sample consists of all galaxies from the $z=0$ snapshot of the TNG50 simulation with stellar mass\footnote{{{All masses in this paper correspond to total gravitationally bound masses (unless noted otherwise).}}} $M_\star > 10^9~{\text{M}}_\odot$ (subhaloes with {\tt{subhaloFlag}} = 0 are discarded, as these are considered not to have a cosmological origin). This lower limit on stellar mass is imposed to ensure that the simulated galaxies have sufficient resolution (the baryonic mass resolution of the TNG50 simulation is $8.5\times10^4~{\text{M}}_\odot$). {{This is needed to minimise the effects of spurious collisional heating by dark matter particles \citep{2023MNRAS.519.5942W, 2023MNRAS.525.5614L}.}} To select star-forming galaxies, we applied an additional threshold on the specific star formation rate, ${\text{sSFR}} > 10^{-11}~{\text{yr}}^{-1}$ \citep{2004MNRAS.351.1151B, 2009MNRAS.397.1776F, 2019MNRAS.485.4817D, 2023A&A...669A..11P}. To ensure that the galaxies are bona-fide, disc-dominated galaxies, we imposed the additional requirement $M_{\text{thin}}/M_\star > 0.5$, where $M_{\text{thin}}$ is the stellar mass of the thin disk component. The latter quantity is determined by means of a structural decomposition on stellar kinematics with the MORDOR code \citep{2022MNRAS.515.1524Z} {{and is obtained from the TNG data archive \citep{2019ComAC...6....2N}.}} The number of TNG50 galaxies satisfying these criteria is 925.

For each galaxy, we generate synthetic observations corresponding to five different random observer positions, with four of them to be used in the analysis (see Sec.~{\ref{AbsoluteMagnitudes.sec}}). We consider each of these four datasets as an independent dataset, which enlarges our sample to $4 \times 925 = 3700$ individual datasets (hereafter simply called galaxies). We only use galaxies with inclinations above 45 degrees, to avoid large uncertainties in the inclination correction for the circular velocity. We also discarded galaxies with inclinations above 85 degrees {{to avoid too large dust attenuation uncertainties}}. In total, we end up with 2151 individual galaxies, which we refer to as our TNG50 sample.

Fig.~{\ref{SampleHistograms.fig}} shows histograms of a number of important galaxy characteristics of our TNG50 galaxy sample. We also show the same histograms for the \citetalias{2001ApJ...563..694V} and \citetalias{2017MNRAS.469.2387P} samples, the two observational TFR studies to which we compare our results. The H{\sc{i}} masses of our TNG50 galaxies cover the range between $10^8$ and $10^{10.5}~{\text{M}}_\odot$, with a peak in the distribution around $10^{9.5}~{\text{M}}_\odot$. The stellar masses range between $10^9$ and $10^{11.5}~{\text{M}}_\odot$, with a maximum in the distribution just below $10^{10}~{\text{M}}_\odot$. Note that this distribution is quite different from the stellar mass distribution of the entire TNG50 galaxy population, which keeps increasing towards smaller stellar masses. Our selection criteria on sSFR and $M_{\text{thin}}/M_\star$ are responsible for the removal of many low-mass galaxies from the sample. The distribution in sSFR ranges from $10^{-11}$ to $10^{-9.5}~{\text{yr}}^{-1}$ with a strong peak in the distribution around $10^{-10}~{\text{yr}}^{-1}$.  Finally, the inclinations of both samples are nicely distributed between 45 and 85 deg, with a slight preference for the highest inclinations, as expected for a random distribution in viewing angle. In general, the distribution of the physical properties of the simulated galaxies are in fair agreement with the characteristics of the galaxies of the \citetalias{2001ApJ...563..694V} and \citetalias{2017MNRAS.469.2387P} samples, even though they have been obtained using different methods (see caption of Fig.~{\ref{SampleHistograms.fig}} for more details).

To further investigate whether our TNG50 galaxy sample is representative of the entire spiral galaxy population in the Local Universe, we show two scatter plots in Fig.~{\ref{RESOLVE.fig}}. Both panels contain, apart from the simulated galaxies from our TNG50 galaxies and the galaxies from the \citetalias{2001ApJ...563..694V} and \citetalias{2017MNRAS.469.2387P} samples, a large set of galaxies from the REsolved Spectroscopy of a Local VolumE (RESOLVE) survey, a volume-limited census of stars and gas in the nearby universe \citep{2015ApJ...810..166E, 2016ApJ...832..126S}. The RESOLVE sample spans different environments and is complete down to baryonic masses of about $10^{9.2}~{\text{M}}_\odot$. The left and right panels show the ${\textit{g}}-{\textit{r}}$ colour and the H{\sc{i}} mass as a function of the ${\textit{K}}_{\text{s}}$-band absolute magnitude, respectively. The four populations generally agree very well in these plots. One difference is that the RESOLVE sample contains a population of redder objects than our TNG50 sample, at all ${\textit{K}}_{\text{s}}$-band absolute magnitude. The presence of these red sequence galaxies is is not surprising, since the RESOLVE sample contains all types of galaxies from different environments, whereas our TNG50 sample only contains star-forming disc galaxies. We note that also the \citetalias{2017MNRAS.469.2387P} sample contains three luminous galaxies with very red ${\textit{g}}-{\textit{r}}$ colours, clearly located in the red sequence region. Two out of these three galaxies are early-type spiral galaxies with sSFR values below the $10^{-11}~{\text{yr}}^{-1}$ threshold, while the other one is a highly inclined and strongly reddened spiral galaxy.

Overall, based on Figs.~{\ref{SampleHistograms.fig}} and~{\ref{RESOLVE.fig}}, we argue that our sample is representative for the population of disc-dominated, late-type galaxies in the local Universe.

%%%%%%%%%%%%%%%%%%%%%%%%%%%%%%%%%%%%%%%%%%%%%%%%%%%%%%

\section{Data generation}
\label{DataGeneration.sec}

The TFR is a correlation between the total luminosity or absolute magnitude in optical/NIR wavebands and the inclination-corrected width of the H{\sc{i}} line profile for disk galaxies. In this section we describe how we determine these quantities for the simulated galaxies in our TNG50 sample. 

\subsection{Multi-wavelength absolute magnitudes}
\label{AbsoluteMagnitudes.sec}

We used the SKIRT code \citep{2011ApJS..196...22B, 2015A&C.....9...20C, 2020A&C....3100381C} to generate broadband fluxes for each of the galaxies in our sample. SKIRT is a general-purpose Monte Carlo radiative transfer code that can be used for dust radiative transfer \citep{2003MNRAS.343.1081B, 2011ApJS..196...22B}, Ly$\alpha$ resonant line scattering \citep{2021ApJ...916...39C}, X-ray radiative transfer \citep{2023A&A...674A.123V, 2024A&A...689A.297V}, and atomic and molecular line radiative transfer \citep{2023A&A...678A.175M}. Its prime application is the generation of synthetic data for simulated galaxies extracted from cosmological hydrodynamical simulations \citep[recent examples include][]{2024arXiv240714061B, 2024MNRAS.527.6506B, 2024MNRAS.530.3765G, 2024MNRAS.531.3839G, 2024arXiv240811037P, 2024ApJ...971..111R, 2024MNRAS.tmp.2560F, 2024arXiv241108117J}. The code accounts for the absorption, scattering, and polarisation by interstellar dust, incorporating the different stellar populations and the complex star--dust geometry. 

To generate the absolute magnitudes, {{we use SKIRT in its {\tt{ExtinctionOnly}} mode, which avoids the costly calculation of the thermal dust emission \citep{2015A&A...580A..87C}.}} We largely follow the modelling strategy described by \citet{2024A&A...683A.181B}, which builds on previous work by \citet{2016MNRAS.462.1057C, 2018ApJS..234...20C, 2022MNRAS.512.2728C}, \citet{2017MNRAS.470..771T}, \citet{2021MNRAS.506.5703K}, and \citet{2022MNRAS.516.3728T}. We start by selecting from the TNG archive \citep{2019ComAC...6....2N} all the stellar particles and gas cells belonging to each simulated galaxy in our sample. We assign a template spectrum to each stellar particle. Particles older than 30~Myr are assigned a simple stellar population spectrum from the BPASS template library \citep{2017PASA...34...58E}, whereas particles younger than 30~Myr are assigned a star-forming region template from the new TODDLERS library \citep{2023MNRAS.526.3871K, 2024A&A...692A..79K}. Photon packets are randomly generated from these sources and are propagated through the interstellar medium (ISM), where they are subject to absorption and scattering by dust grains. The dusty medium is discretised onto an octree with up to 12 levels of refinement \citep{2013A&A...554A..10S, 2014A&A...561A..77S}. Interstellar dust is not followed as a separate species in the TNG50 simulation, but is added in post-processing to the ISM assuming a constant dust-to-metal ratio $f_{\text{dust}} = 0.2$ \citep{2022MNRAS.516.3728T}, a value determined by calibrating TNG50 fluxes to observed galaxies from the DustPedia sample \citep{2017PASP..129d4102D, 2018A&A...609A..37C}. To discriminate {{between the cool dusty ISM and the hotter dust-free circumgalactic medium}}, we use the prescription by \citet{2012MNRAS.427.2224T, 2019MNRAS.484.5587T}, which defines a separation based on gas temperature and density. We use the diffuse ISM THEMIS model \citep{2017A&A...602A..46J} as our dust grain model. 

{{SKIRT allows}} an arbitrary number of observer positions and different options for the instruments, including broadband filter photometers \citep{2008MNRAS.391..617B, 2020A&C....3100381C}. Following the setup for the TNG50-SKIRT Atlas \citep{2024A&A...683A.181B}, we choose five observers for each galaxy, spread on the unit sphere in optimal arrangement. Their position is chosen relative to the simulation box, which has no connection to the orientation of each individual galaxy. Two of the observer positions (O4 and O5) are antipodal, resulting in a semi-redundancy. We used the data corresponding to the O4 observer in the analysis and used the O5 position for consistency checks \citep[see also][]{2024A&A...683A.182B}. Details of the observer positions can be found in Table 1 of \citet{2024A&A...683A.181B}. For each observer position, we calculate fluxes in the 12 UV to NIR broadband filters considered by \citetalias{2017MNRAS.469.2387P}: the {\em{GALEX}} FUV and NUV bands, the SDSS {\em{ugriz}} bands, the 2MASS {\em{JHK}}$_{\text{s}}$ bands, and the {\em{Spitzer}} IRAC [3.6] and [4.5] bands. 

An important aspect to consider is the correction for internal dust attenuation. We mimic the observational approach as applied in TFR studies and apply a correction on the dust-attenuated fluxes. Two different schemes are commonly used in TFR studies \citep[for a discussion, see][]{2001ApJ...563..694V}. The first scheme, first presented by \citet{1998AJ....115.2264T} and refined by \citet{2019ApJ...884...82K}, is empirical in nature. In each band, the parameters of the attenuation, which can depend on a number of galaxy parameters, are chosen to minimise the scatter in the TFR. The second scheme, proposed by \citet{1985ApJS...58...67T} and adapted by \citet{1996AJ....112.2471T}, is a physically motivated scheme based on an analytical two-phase radiative transfer model. {{For our simulated galaxies, we}} opt for the second attenuation correction method. {{We come back to the dust attenuation correction in Sec.~{\ref{DustAttenuation.sec}}.}}

{{The final step consists of converting the dust-attenuation-corrected fluxes to absolute magnitudes using the standard equations (throughout this paper we use AB magnitudes). Following \citetalias{2020ApJ...896....3K}, we assign conservative uncertainties of 0.10~mag to the FUV, NUV and {\it{u}}-band absolute magnitudes, and 0.05~mag to the absolute magnitudes in the other bands.}}

\subsection{HI line widths}
\label{LineWidths.sec}

Apart from the multi-band absolute magnitudes, we need to calculate the synthetic H{\sc{i}} line profile and calculate its width for each observer position and for each galaxy in our final sample. In the TNG50 simulation, the cold gas is not automatically split in its atomic and molecular contributions {{since TNG50 just doesn't model the cold gas (below $\sim10^4$~K) since the relevant chemistry and cooling processes are not implemented.}} This H{\sc{i}}-H$_2$ partitioning thus needs to be done in post-processing using subgrid recipes \citep{2015MNRAS.452.3815L, 2018ApJS..238...33D}. 

Various H{\sc{i}}-H$_2$ partitioning recipes are available in the literature, with different levels of sophistication \citep[e.g.,][]{2006ApJ...650..933B, 2008AJ....136.2782L, 2011ApJ...728...88G, 2013MNRAS.436.2747K, 2014ApJ...795...37G, 2014ApJ...790...10S, 2024ApJ...966..172P}. Our approach is based on the framework described by \citet{2023MNRAS.521.5645G}. {{We first run a SKIRT radiative transfer simulation in {\tt{ExtinctionOnly}} mode}} to determine, in each individual cell in the ISM, the radiation field strength at 1000~\AA, a commonly adopted proxy for the radiation field strength in the Lyman--Werner band. This first step largely follows the steps discussed in the previous subsection, with the exception that we store the radiation field strength in every individual cell. Once this first step is finished, we use the UV radiation field, together with the local gas density, the metallicity and the gas cell size, to determine the fraction of H{\sc{i}} in each cell. This is achieved using the subgrid recipe of \citet{2014ApJ...795...37G}, which is an update of \citet{2011ApJ...728...88G}. With the H{\sc{i}} density in each cell determined, we run a second SKIRT simulation, {{now in the {\tt{GasEmission}} mode,}} with the H{\sc{i}} gas emitting the photon packets. These photon packets are propagated to the observers, accounting for the Doppler shifts and thermal broadening as described by \citet{2020A&C....3100381C}. 

\begin{figure*}
\centering
\includegraphics[width=\textwidth]{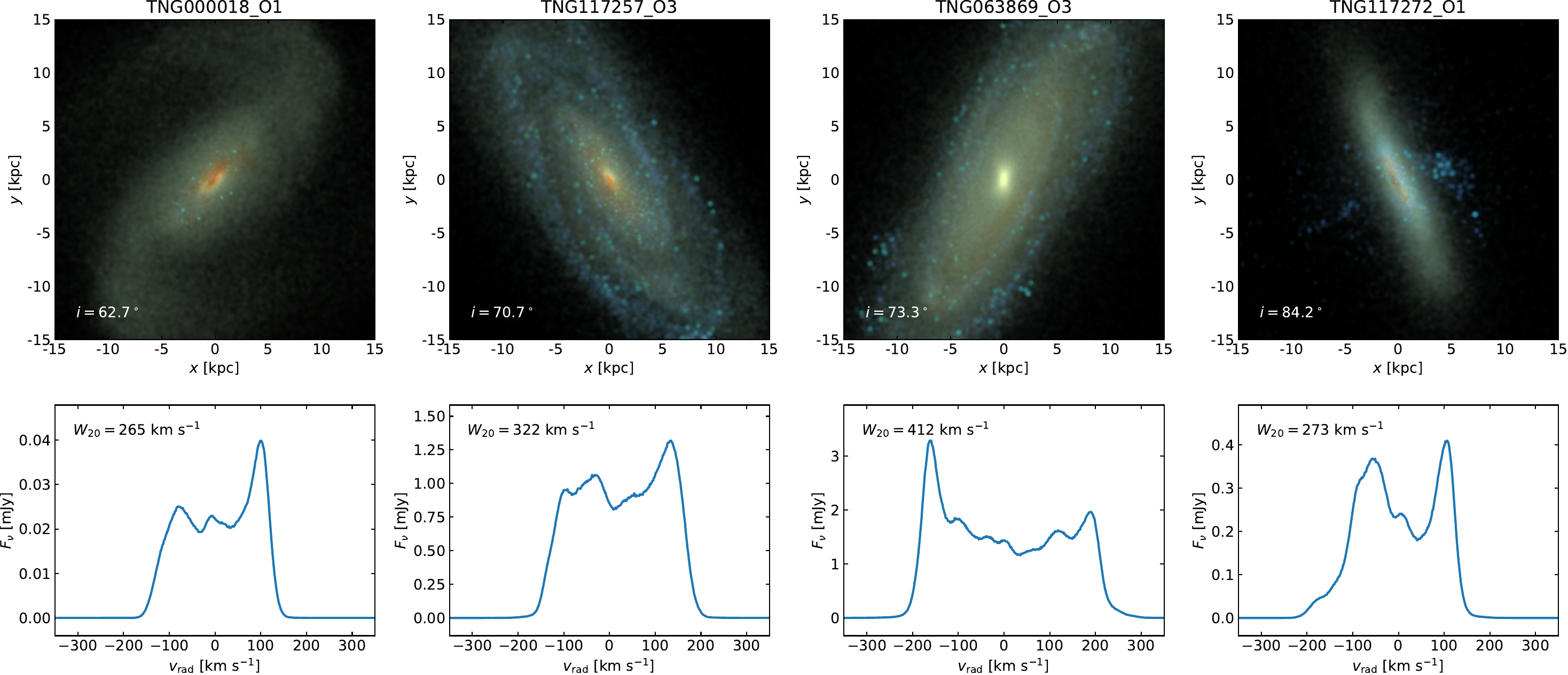}
\caption{Three-colour images (top row) and H{\sc{i}} profiles (bottom row) for four galaxies from our sample. The images have a field-of-view of 30 kpc, and combine the SDSS ${\textit{g}}$, ${\textit{r}}$ and ${\textit{z}}$ band images according to the methodology of \citet{2004PASP..116..133L}. The inclinations and the $W_{20}$ line widths are indicated.}
\label{visual.fig}
\end{figure*}

The final result is a H{\sc{i}} line profile for every observer. We have chosen a velocity resolution of 5~km\,s$^{-1}$ in our synthetic detectors. The H{\sc{i}} line width $W_{20}$ is determined as the width of the total profile at 20\% of the peak value. We note that different approaches are adopted in the literature to determine the H{\sc{i}} line width. Some studies, including the original TFR study \citep{1977A&A....54..661T},  adopt the width of the total H{\sc{i}} line profile at 20\% of the peak level, an approach we follow. Other studies, including \citetalias{2001ApJ...563..694V} apply a more complex methodology in which, in case of a double-peaked profile, the peak fluxes on both sides of the profile are considered separately for the calculation of the 20\% level \citep{2001A&A...370..765V}. Other studies consider the line width at 50\% of the peak or average H{\sc{i}} flux, or use more complex algorithms that are optimised for noisy H{\sc{i}} profiles (\citealt{1997AJ....113...53G, 2005ApJS..160..149S, 2014MNRAS.443.1044M}; \citetalias{2017MNRAS.469.2387P}; \citetalias{2020ApJ...896....3K}). Beside line widths based on the integrated H{\sc{i}} profile, one can also consider characteristic velocities based on resolved rotation curves, such as the maximum rotation velocity or the amplitude of the rotation curve in the outer flat part \citepalias[e.g.,][]{2001ApJ...563..694V, 2017MNRAS.469.2387P}. We return to the issue of the choice of the characteristic velocity scale in Sec.~{\ref{Towards.sec}}.

The line width is corrected for the inclination, $W_{20}^i = W_{20}/\sin i$, where $i$ represents the inclination. While the inclination for observed galaxies is typically determined from axis ratios in optical images or from spatially resolved kinematics, we determined $i$ as the angle between the line of sight and the angular momentum vector of the stellar particles within twice the stellar half-mass radius \citep{2018ApJS..238...33D}. We will discuss this aspect more in Sec.~{\ref{Towards.sec}}. We assign a random uncertainty $\Delta \log W_{20}^i$ to each inclination-corrected line width by sampling from a Gaussian distribution with mean 0.027 and standard deviation 0.017, based on measurements from \citetalias{2020ApJ...896....3K}.

In Fig.~{\ref{visual.fig}} we show synthetic RGB images and H{\sc{i}} line profiles for four representative simulated galaxies of our TNG50 sample.

%%%%%%%%%%%%%%%%%%%%%%%%%%%%%%%%%%%%%%%%%%%%%%%%%%%%%%

\section{Results}
\label{TNG50-TF.sec}

\subsection{The TNG50 multi-wavelength TFR}

\begin{figure*}
\includegraphics[width=\textwidth]{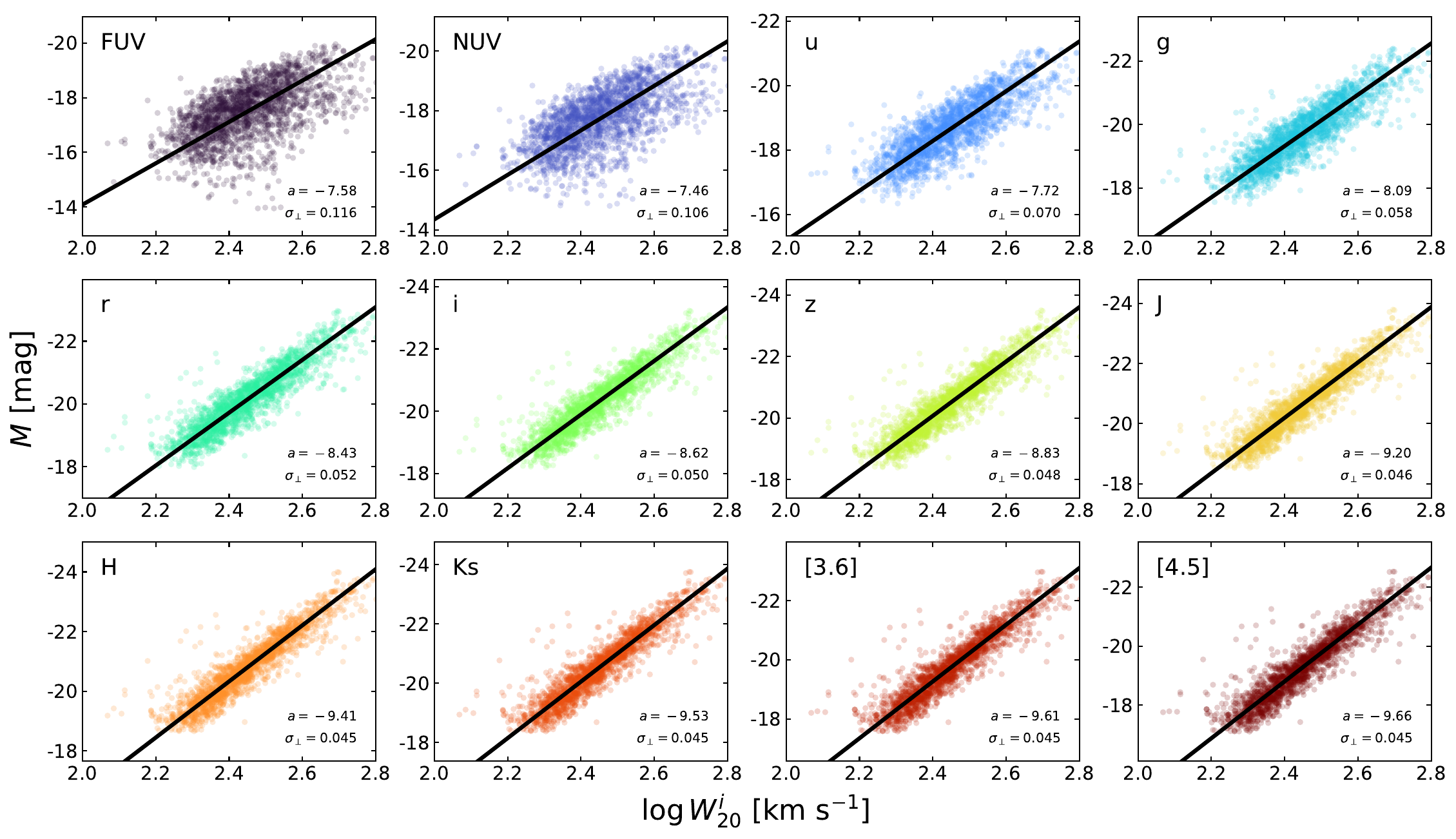}
\caption{The multi-wavelength TFR for the simulated disk galaxies in our TNG50 sample. The different panels correspond to different broadband filters in order of increasing wavelength. The slope and the tightness of the TFR are indicated in the bottom right corner of each panel.}
\label{TNG50-TF.fig}
\end{figure*}

Fig.~{\ref{TNG50-TF.fig}} shows the main result of this paper: the multi-wavelength TFR for the 12 broadband filters we consider. In every band, there is a clear correlation between the absolute magnitude and the inclination-corrected H{\sc{i}} line width. It is obvious, even by eye, that the correlation can be well approximated by a power-law, and that the scatter decreases with increasing wavelength. Indeed, the relation has significant scatter in the UV bands and becomes tighter when we move towards optical, NIR and MIR bands.

To quantify the TFR, we apply linear regression, that is, we fit a straight line to the data of the form
\begin{equation}
M = a \left( \log W_{20}^i - \langle \log W_{20}^i\rangle \right) + b,
\end{equation}
with $a$ the slope of the line, and $b$ the intercept at the pivot value $\langle \log W_{20}^i\rangle$, the mean value of $\log W_{20}^i$ for the galaxies in our sample \citep{2016A&A...593A..39P}. This is a trivial textbook problem in the perfect-world case of no intrinsic scatter and Gaussian distributed noise of known amplitudes in the $y$ direction only. The problem becomes significantly more cumbersome in the presence of intrinsic scatter, correlated and/or non-Gaussian uncertainties in both quantities, and bad data (outliers). This has resulted in a plethora of more advanced linear regression methods \citep[e.g.,][]{JogeshBabu1992, 1996ApJ...470..706A, 2004AmJPh..72..367Y, 2010arXiv1008.4686H, Saracli2013, 2018AMT....11.1233W, Jing2024}. 

Following \citetalias{2017MNRAS.469.2387P}, we use the Python code BCES \citep{2012Sci...338.1445N}, which implements the robust bivariate linear regression technique presented by \citet{1996ApJ...470..706A}. The black lines in Fig.~{\ref{TNG50-TF.fig}} are the result of this orthogonal regression method. The details of the fit are listed in Table~{\ref{TNG50-TF.tab}}. Apart from the slope and the zero-point, we also list the tightness of the fit. As its symbol $\sigma_\perp$ indicates, it is defined as the perpendicular scatter\footnote{{{Note that the name `tightness' for $\sigma_\perp$ is slightly counter-intuitive: it refers to the level of scatter, so the $\sigma_\perp$ decreases as the relation becomes tighter. To avoid confusion: we use the terminology that the tightness improves if the relation becomes tighter and $\sigma_\perp$ decreases.}}} between the data points and the best fitting linear model (for details, see \citealt{2016A&A...593A..39P}; \citetalias{2017MNRAS.469.2387P}). 

\begin{table}
\caption{Slope, zero-point, and tightness of the multi-wavelength TFR of our TNG50 galaxy sample.}
\label{TNG50-TF.tab}
\centering
\begin{tabular}{cccc}
band & slope $a$ & zero-point $b$ & tightness $\sigma_\perp$ \\
 & [mag\,dex$^{-1}$] & [mag] & [dex] \\[0.5em]
\hline \\
FUV				& $-7.575 \pm 0.160$ 	& $-17.429 \pm 0.019$ 	& $0.116 \pm 0.006$ \\	
NUV				& $-7.464 \pm 0.142$ 	& $-17.710 \pm 0.017$ 	& $0.106 \pm 0.005$ \\	
{\it{u}}			& $-7.717 \pm 0.103$ 	& $-18.666 \pm 0.012$ 	& $0.070 \pm 0.002$ \\	
{\it{g}}			& $-8.089 \pm 0.092$ 	& $-19.728 \pm 0.010$ 	& $0.058 \pm 0.001$ \\
{\it{r}}			& $-8.427 \pm 0.089$ 	& $-20.141 \pm 0.010$ 	& $0.052 \pm 0.001$ \\	
{\it{i}}			& $-8.618 \pm 0.087$ 	& $-20.323 \pm 0.009$ 	& $0.050 \pm 0.001$ \\	
{\it{z}}			& $-8.833 \pm 0.087$ 	& $-20.510 \pm 0.009$ 	& $0.048 \pm 0.001$ \\	
{\it{J}}			& $-9.201 \pm 0.088$ 	& $-20.654 \pm 0.009$ 	& $0.046 \pm 0.001$\\	
{\it{H}}			& $-9.407 \pm 0.089$ 	& $-20.799 \pm 0.009$ 	& $0.045 \pm 0.001$ \\	
{\it{K}}$_{\text{s}}$	& $-9.530 \pm 0.089$ 	& $-20.530 \pm 0.009$ 	& $0.045 \pm 0.001$ \\	
$[3.6]$			& $-9.607 \pm 0.090$ 	& $-19.745 \pm 0.009$ 	& $0.045 \pm 0.001$ \\	
$[4.5]$			& $-9.662 \pm 0.091$ 	& $-19.284 \pm 0.009$ 	& $0.045 \pm 0.001$ \\[0.5em]
\hline	
\end{tabular}
\end{table}

Fig.~{\ref{SlopeTightness.fig}} shows how the slope (top panel) and the scatter (bottom panel) of the multi-wavelength TFR relation change with increasing wavelength. Neglecting the FUV band, the slope varies systematically with wavelength over the entire UV to MIR wavelength range, in the sense that the TFR gradually steepens. The shallowest slope of $-7.46\pm0.14~{\text{mag}}~{\text{dex}}^{-1}$ is obtained in the NUV band, and the relation steepens towards a steep slope of $-9.66\pm0.09~{\text{mag}}~{\text{dex}}^{-1}$ in the IRAC [4.5] band. The tightness of the TNG50 TFR also shows a prominent and systematic trend as a function of wavelength: it starts at $0.116\pm0.006$~dex in the FUV band and {{improves systematically}} in the optical regime until it reaches the value of $0.046\pm0.001$~dex in the {\it{J}} band. It remains nearly constant at this level over the entire NIR and MIR regime. 

\subsection{Comparison to observations}

\begin{figure}
\includegraphics[width=0.95\columnwidth]{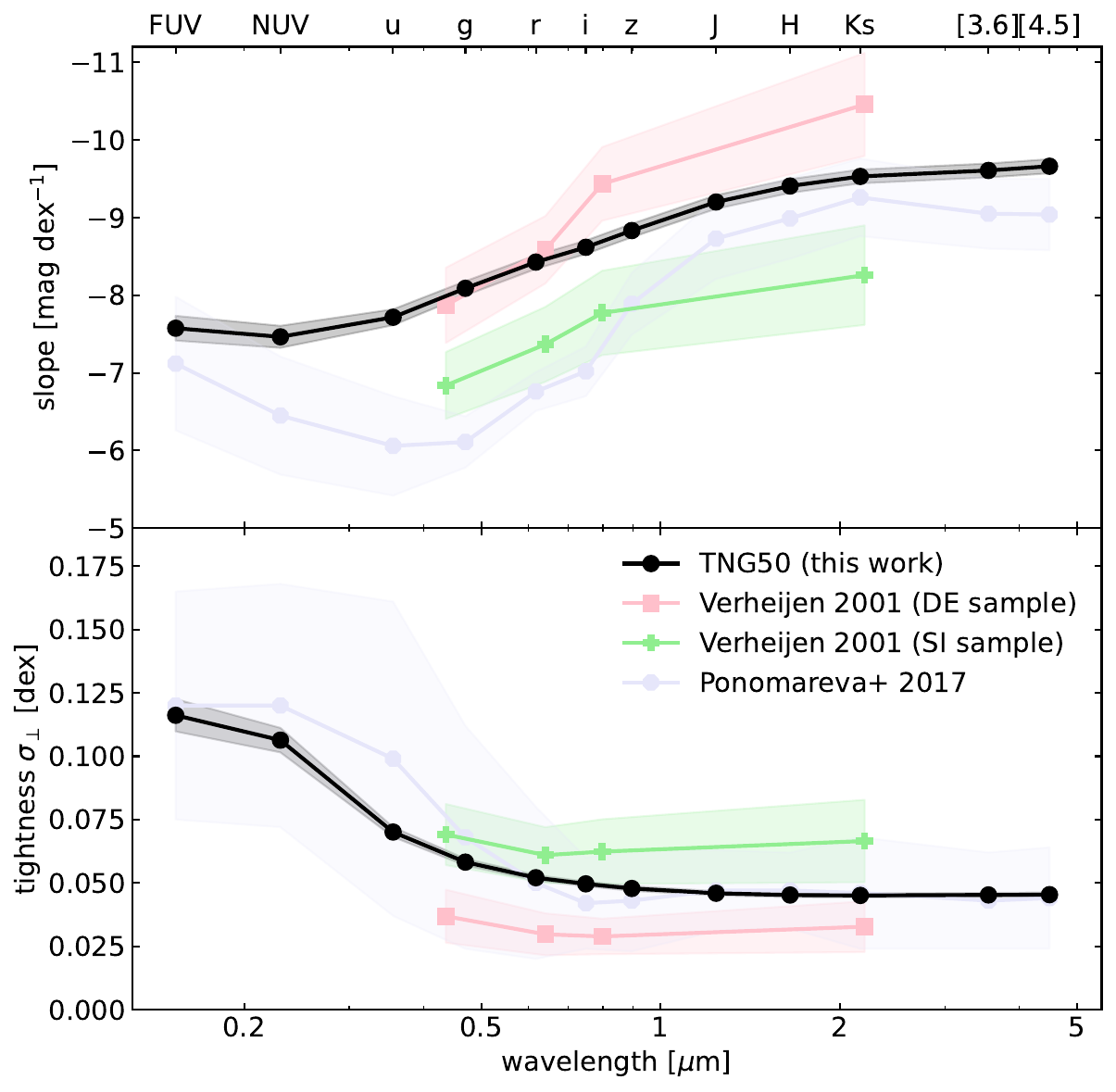}%
\caption{Slope (top panel) and tightness (bottom) panel of the TFR as a function of wavelength. Black dots are our TNG50 TFR results, the coloured lines are the results from the observational samples of \citetalias{2001ApJ...563..694V} and \citetalias{2017MNRAS.469.2387P}. All results in this figure are based on the same fitting method. {{Shaded regions indicate 1-$\sigma$ uncertainty intervals.}}}
\label{SlopeTightness.fig}
\end{figure}

\begin{figure*}
\centering
\includegraphics[width=0.8\textwidth]{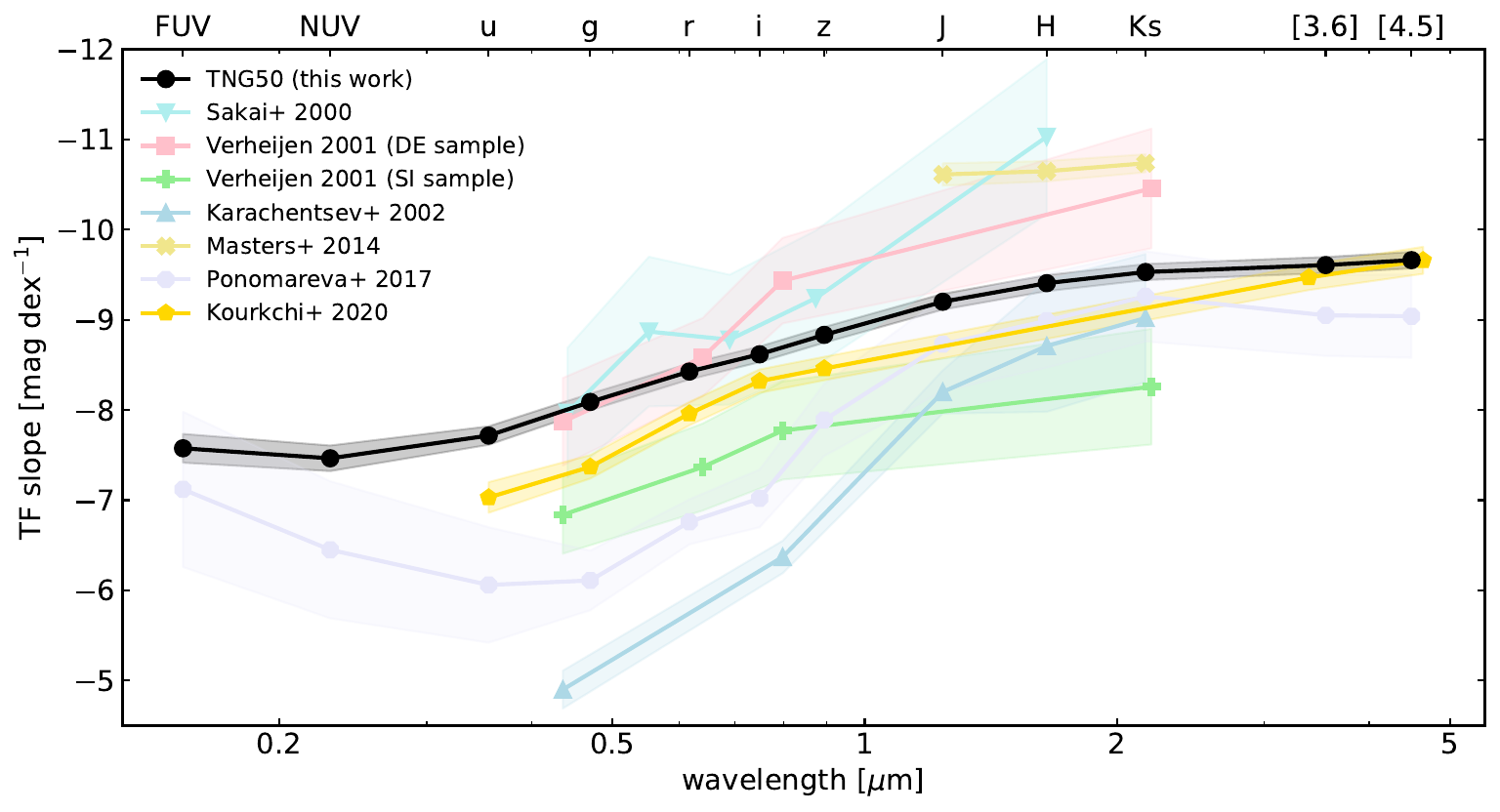}
\caption{Comparison of the wavelength dependence of the slope of our TNG50 TFR to the slopes measured by a number of observed multi-wavelength TFRs, as indicated in the top left corner. {{Shaded regions indicate 1-$\sigma$ uncertainty intervals.}} {{Contrary to Fig.~{\ref{SlopeTightness.fig}}, this figure includes a wider selection of observed TFR studies with a more diverse range of fitting strategies.}}}
\label{SlopeComparison.fig}
\end{figure*}

For our comparison to observational TFR studies, we use a two-tiered approach. We first compare our results to the multi-wavelength TFR studies of \citetalias{2001ApJ...563..694V} and \citetalias{2017MNRAS.469.2387P}, based on exactly the same fitting method as ours. Subsequently we relax this condition and we compare our results to a broader selection of observational multi-wavelength TFR studies.

\citetalias{2001ApJ...563..694V} performed a detailed TFR study in the {\it{BRIK'}} bands for spiral galaxies from the Ursa Major Cluster. He considered different subsamples based on different criteria, and he used three different kinematic measures: the width of the global H{\sc{i}} profile, the maximum rotation velocity obtained from the rotation curve, and the velocity in the flat part of the rotation curve. In our comparison, we focused on the results based on the integrated H{\sc{i}} profiles. In order to eliminate as many biases as possible, we have fitted the TFR to the data that he presents using the same fitting routine as we have used for our analysis rather than directly using the fitting results from his study. From the different options presented, we have selected two representative samples: the SI sample (38 galaxies for which useful H{\sc{i}} synthesis imaging is available) and for the more restrictive DE sample (12 non-interactive, late-type, unbarred galaxies with smooth outer isophotes). In both cases, the TFR slopes we have calculated are compatible with the slopes published by \citetalias{2001ApJ...563..694V} within the error bars. 

\citetalias{2017MNRAS.469.2387P} constructed the multi-wavelength TFR for a sample of 32 galaxies, in the same 12 UV-to-MIR broadband filters that we consider in our study. Contrary to some other TFR studies which aim at maximising the sample size, they aimed at constructing a high-quality TFR study of a modest but representative sample of galaxies with independent distance measurements from the Cepheid period-luminosity relation and/or TRGB stars. All galaxies in their sample have interferometric H{\sc{i}} data and global H{\sc{i}} profiles without evidence of distortion or blending. Following \citetalias{2001ApJ...563..694V}, they also use different kinematic measures, and again, we compare our results to the results they obtained from the integrated H{\sc{i}} profiles. The results were obtained {{with the same fitting methodology}} as we apply for our TNG50 sample.

The two panels of Fig.~{\ref{SlopeTightness.fig}} show, apart from our TNG50 results, the wavelength dependence of the slope and the tightness of {{the TFR from}} \citetalias{2001ApJ...563..694V} (separately for the SI and DE samples) and \citetalias{2017MNRAS.469.2387P}. For the three different samples, we see the same trend as for our TNG50 data: as we move to longer wavelengths, the TFR becomes steeper and tighter. Looking more quantitatively at the wavelength dependence of the slope and tightness, a number of aspects are worth noting. 

Interestingly, the slope of our TFR seems to be sandwiched between the values of the two \citetalias{2001ApJ...563..694V} subsample slopes: the SI slope is consistently shallower than ours, whereas the DE slope is consistently steeper. At the same time, the DE TFR is tighter than our TNG50 TFR, which in turn is tighter than the SI TFR. It is clear that the details of the sample selection play a major role for the characteristics of the TFR, an aspect to which we return in Sec.~{\ref{Towards.sec}}. 

Comparing the slope of our TNG50 sample to the slope determined by \citetalias{2017MNRAS.469.2387P}, we note that our slopes systematically steepen from the NUV to the [4.5] band, whereas their slopes only steepen over a slightly smaller wavelength interval, from the {\it{u}} band to the {\it{K}}$_{\text{s}}$ band. The most important difference is that they found a significantly shallower TFR in the blue optical bands. The difference is largest in the {\it{g}} band: while we find a slope of $-8.11\pm0.10~{\text{mag}}~{\text{dex}}^{-1}$, they found the much shallower value of $-6.11\pm0.33~{\text{mag}}~{\text{dex}}^{-1}$. In the near-infrared bands, the slopes agree quite well. Turning to the bottom panel, the agreement between the tightness of our TNG50 TFR and the one obtained by \citetalias{2017MNRAS.469.2387P} is striking. 

In Fig.~{\ref{SlopeComparison.fig}} we extend the comparison of the wavelength dependence of the TFR slope to a wider selection of multi-wavelength TFR studies, now not necessarily with a similar fitting strategy. Apart from our own results and the \citetalias{2001ApJ...563..694V} and \citetalias{2017MNRAS.469.2387P} results already discussed, this figure contains the TFR slopes obtained by \citet{2000ApJ...529..698S}, \citet{2002A&A...396..431K}, \citet{2008AJ....135.1738M, 2014AJ....147..124M}, and \citet{2020ApJ...896....3K}. From each of these studies, if several versions of the TFR were available, we have selected the one that was most similar to our approach.

\citet{2000ApJ...529..698S} derived the TFR in the {\it{BVRIH}} bands based on a sample of 21 galaxies with Cepheid distances. In the optical, their slopes agree very well with our determinations, whereas the slope they measure in the {\it{H}} band is particularly steep. \citet{2002A&A...396..431K} considered a sample of 450 galaxies selected from the Revised Flat Galaxy Catalog \citep{1999BSAO...47....5K}. They find very flat slopes, which is probably due to the fact that they selected galaxies that are very close to edge-on, which significant attenuation uncertainties. \citet{2008AJ....135.1738M, 2014AJ....147..124M} measured the TFR in the 2MASS NIR bands based on 888 galaxies selected from the 2MASS Extended Source Catalog, and find a relatively steep TFR when they consider the entire sample. Finally, \citet{2020ApJ...896....3K} present an impressive study of the TFR in the SDSS {\it{ugriz}} and WISE W1 and W2 bands, based on a sample of about 600 galaxies in 20 galaxy clusters in the frame of the Cosmicflows-4 project \citep{2023ApJ...944...94T}. Their slopes agree very well with the ones we have obtained for our TNG50 sample. 

In general, all of these studies consistently show a systematic steepening of the slope with increasing wavelength over the optical--NIR wavelength range. Given the spread among the different studies, we can state that our TNG50 sample reproduces the observed slope of the TFR in a very satisfactory way.

\subsection{Search for a second parameter}

\begin{figure}
\includegraphics[width=0.95\columnwidth]{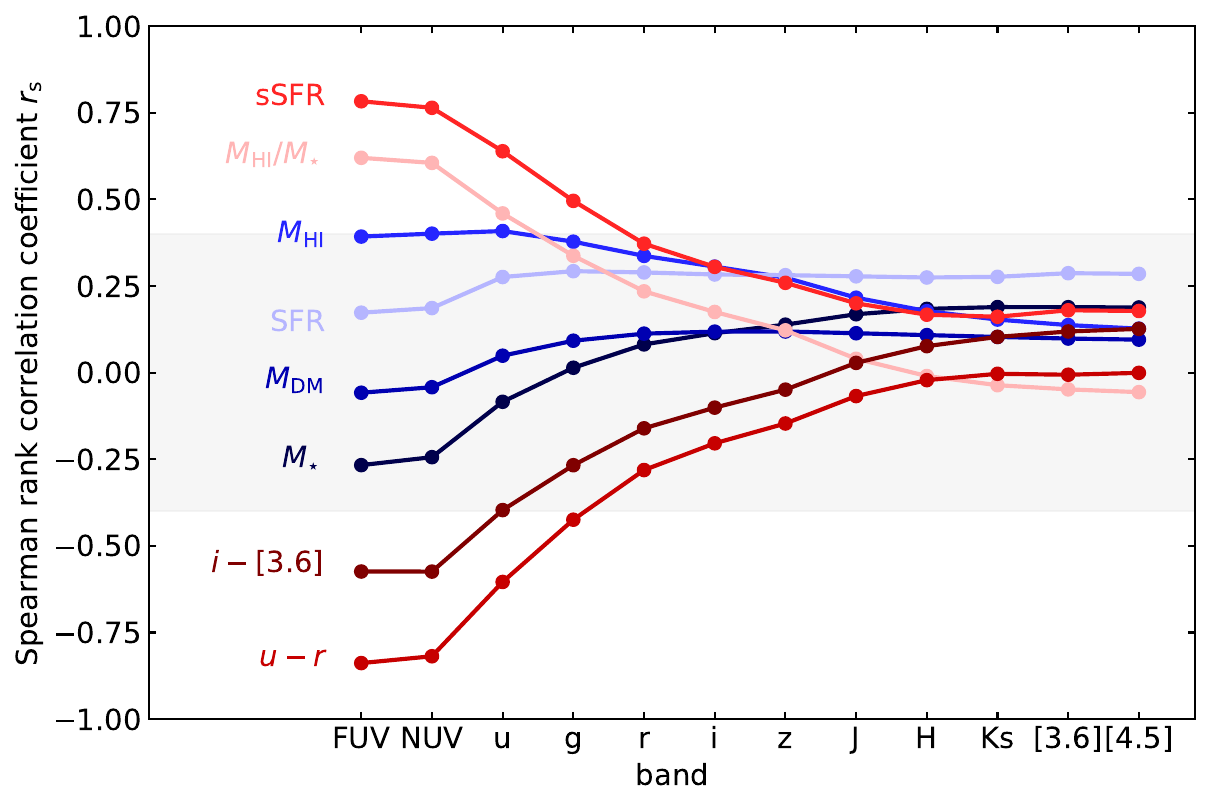}
\caption{Spearman rank correlation coefficients for the correlations between $\Delta M$, the residual between the actual absolute magnitude and the magnitude expected from the TFR, and eight different galaxy properties, as a function of the broadband. Extensive galaxy properties are shown in blue hues, intensive properties in red hues. The grey band corresponds to $|r_{\text{s}}| < 0.4$, usually considered as the region of negligible to weak correlations.}
\label{Spearman.fig}
\end{figure}

\begin{figure*}
\centering
\includegraphics[width=\textwidth]{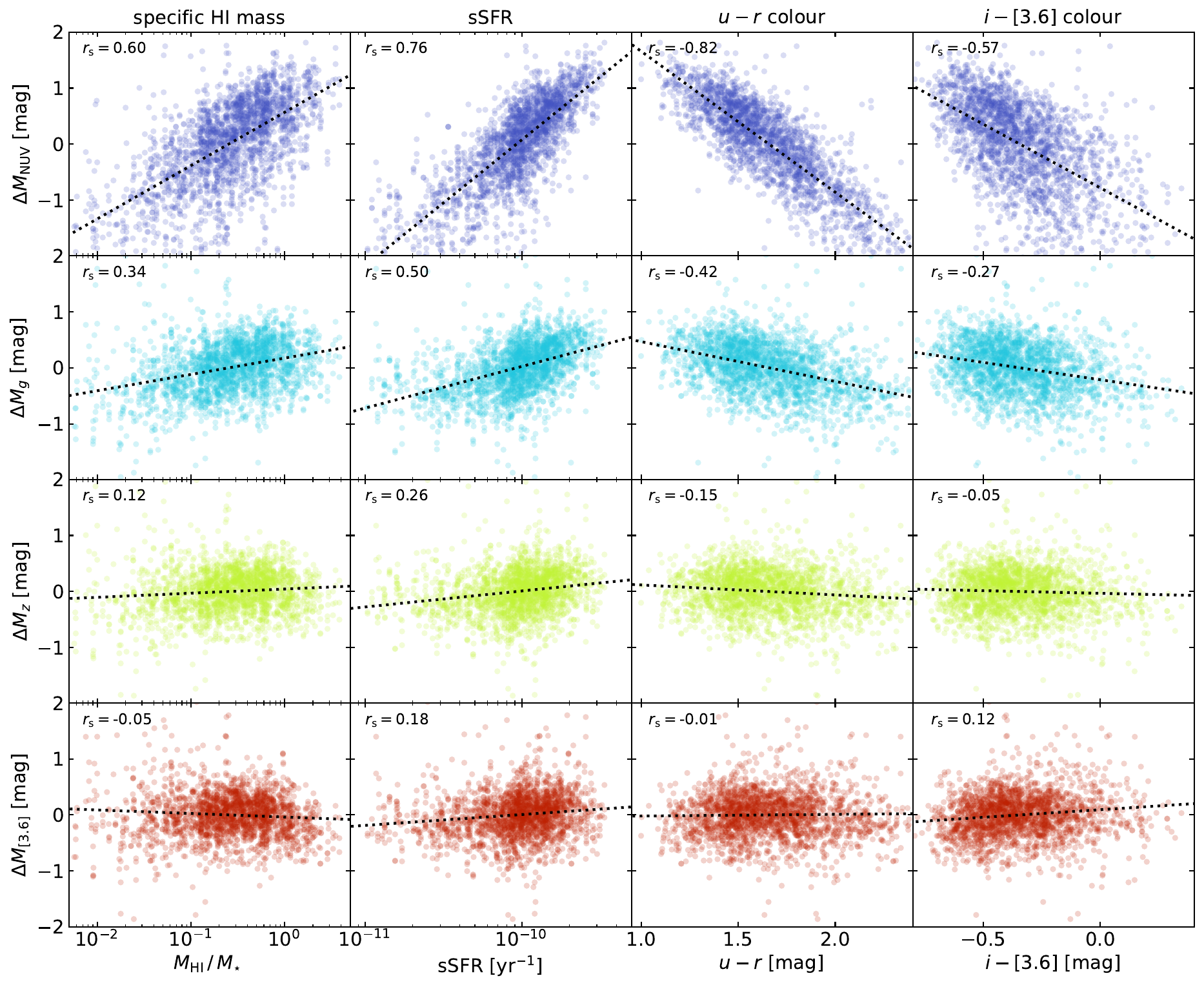}
\caption{Correlations between $\Delta M$, the residual between the actual absolute magnitude and the magnitude expected from the TFR, and four different intensive galaxy properties (from left to right: specific H{\sc{i}} mass, sSFR, ${\textit{u}} - {\textit{r}}$ colour, and ${\textit{i}} - [3.6]$ colour). The different rows correspond to different broadband filters (from top to bottom: NUV, {\textit{g}}, {\textit{z}}, and [3.6]). In each panel, the dotted line is the best-fitting linear fit to the data, and the Spearman rank correlation coefficient is indicated in the top right corner.}
\label{SecondParameter.fig}
\end{figure*}

\begin{figure*}
\includegraphics[width=\textwidth]{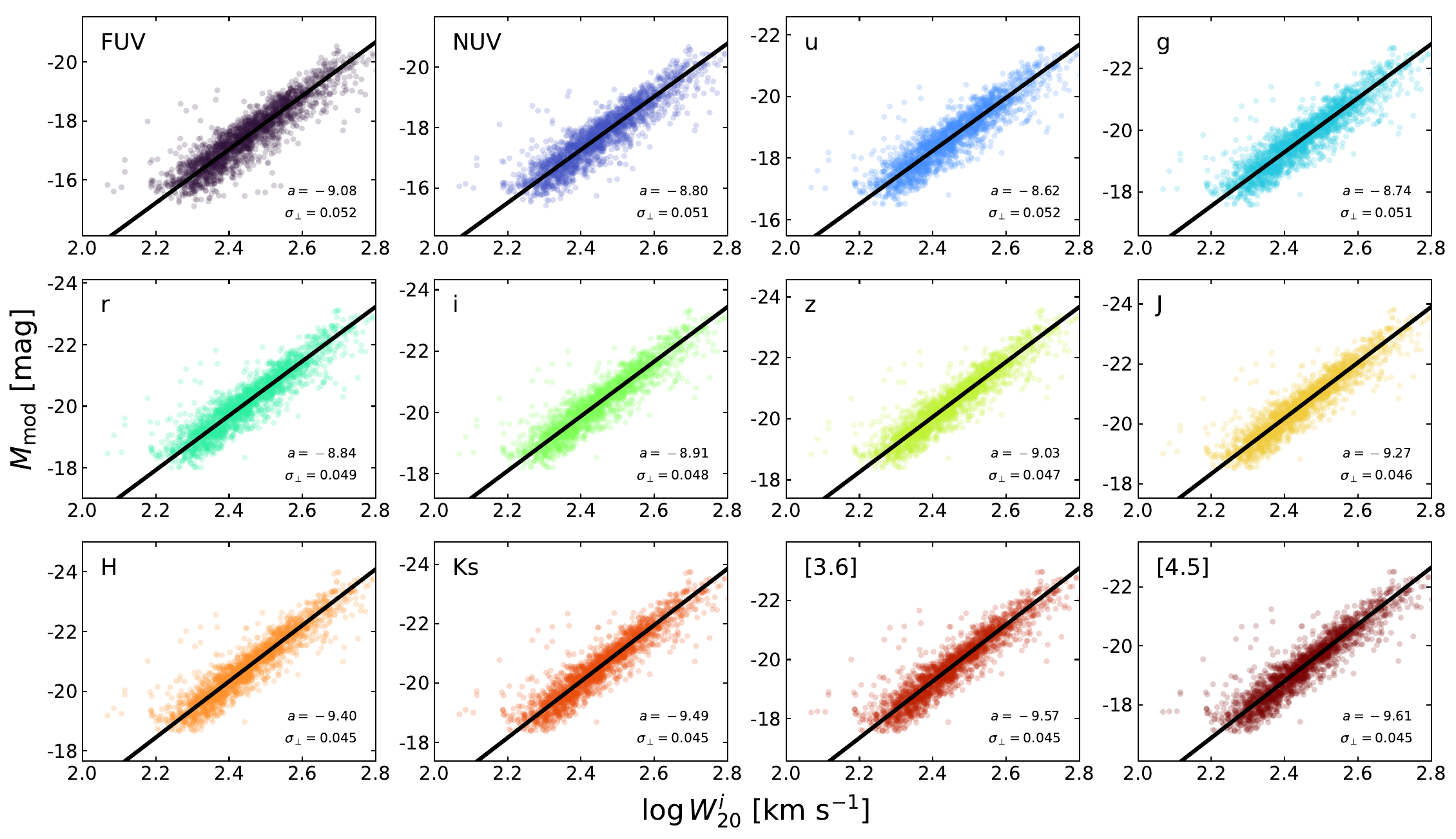}
\caption{The modified multi-wavelength TFR for our TNG50 galaxy sample, with the modified magnitudes based on the ${\textit{u}} - {\textit{r}}$ colour as a second parameter. As in Fig.~{\ref{TNG50-TF.fig}}, the different panels correspond to different broadband filters and the slope and the tightness of the TFR are indicated in the bottom right corner of each panel.}
\label{TNG50-TF-SecondParameter.fig}
\end{figure*}

Many TFR studies have addressed {{the issue of whether}} the scatter in the TFR is purely intrinsic, or whether a second parameter can be called to reduce the scatter. This is usually investigated by correlating the residuals from the best-fitting TFR with other physical properties of the galaxies. Several studies have found that the scatter in the blue bands does correlate with colour or with the galaxy morphological type \citep{1983ApJ...265....1A, 1985ApJ...289...81R, 1997AJ....113...53G, 2004ApJ...607..241R, 2006ApJ...653..861M, 2018MNRAS.479.3373M, 2020ApJ...896....3K}. This correlation tends to weaken towards redder bands. Whether or not a colour term is still capable or reducing the scatter in the near-infrared bands is still matter of discussion: some studies find that the correlations with a colour term disappear in the NIR TFR (\citealt{2001ApJ...563..694V}; \citetalias{2017MNRAS.469.2387P}), others argue that the scatter in the NIR bands can still be reduced by the introduction of a colour adjustment term \citep{2013ApJ...765...94S, 2014ApJ...792..129N}.

As we have access to both synthetic observations and intrinsic physical properties for all of the galaxies in our TNG50 sample, we can easily apply this exercise for our TFR. In each band, we calculate for each galaxy the residual $\Delta M$ between the actual absolute magnitude and the magnitude expected from the TFR. Subsequently we calculate, for a range of physical properties, {{the Spearman rank correlation coefficient}} between $\Delta M$ and the physical property. We explore eight different properties: four extensive properties (stellar mass, dark matter mass, H{\sc{i}} mass, and SFR) and four intensive properties (specific H{\sc{i}} mass, specific SFR, {{dust-corrected}} ${\textit{u}} - {\textit{r}}$ colour, and {{dust-corrected}} ${\textit{i}} - [3.6]$ colour). The Spearman rank correlation coefficients, $r_{\text{s}}$, as a function of wavelength for each of these properties are shown in Fig.~{\ref{Spearman.fig}}. 

Using the conventional interpretation that correlations with $|r_{\text{s}}| < 0.1$ are negligible and correlations with $0.1 < |r_{\text{s}}|< 0.4$ are weak \citep[e.g.,][]{Schober2018}, we find that the correlations between each of the four extensive properties and $\Delta M$ are weak at most (blue hues in Fig.~{\ref{Spearman.fig}}). This even applies to the UV bands in which the scatter of the TFR is quite significant. For the intensive properties (red hues in Fig.~{\ref{Spearman.fig}}) the situation is quite different. For each of the four intensive properties probed, we find a moderate ($0.4 < |r_{\text{s}}| < 0.7$) to strong ($0.7 < |r_{\text{s}}| < 0.9$) correlation with the TFR residuals in the UV bands. The correlations between these intensive properties and the TFR residuals in the NUV, {\textit{g}}, {\textit{z}}, and [3.6] bands are shown in Fig.~{\ref{SecondParameter.fig}}. The strongest correlations are with sSFR or ${\textit{u}} - {\textit{r}}$ colour, two properties that are closely related \citep{2004ApJ...600..681B, 2014MNRAS.440..889S, 2018MNRAS.476...12B}. The strength of each of these correlations quickly decreases for increasing wavelengths, however.  Even the strongest correlations already enter the weak qualification regime in the {\textit{r}} band and they become negligible in the near-infrared bands. For the case of the ${\textit{u}} - {\textit{r}}$ colour, we list the Spearman rank coefficients and $p$-values for the correlation as a function of wavelength in Table~{\ref{ModifiedTFR.tab}}. Based on the $p$-values, the correlation between the residual of the TFR and the ${\textit{u}} - {\textit{r}}$ colour is significant up the ${\textit{J}}$ band.

\begin{table*}
\caption{Second and third column: Spearman rank coefficient, and $p$-value of correlation between the residual $\Delta M$ from the TFR and ${\textit{u}} - {\textit{r}}$ colour. Fourth and fifth column: slope and zero-point of the best-fitting linear between these quantities.}
\label{ModifiedTFR.tab}
\centering
\begin{tabular}{cccrr}
band & $r_{\text{s}}$ & $p$-value & slope\hspace*{2em} & zero-point\hspace*{1em} \\[0.5em]
\hline \\
FUV 				& $-0.839$  & $0$ & $-2.849 \pm 0.038$ 	& $4.725 \pm 0.064$ 	\\
NUV 			& $-0.819$  & $0$ & $-2.518 \pm 0.036$ 	& $4.176 \pm 0.061$ 	\\
{\it{u}}			& $-0.604$  & $0$ & $-1.226 \pm 0.036$ 	& $2.035 \pm 0.060$ 	\\
{\it{g}}			& $-0.425$  & $8.13\times10^{-95}$ & $-0.709 \pm 0.035$ 	& $1.176 \pm 0.060$ 	\\
{\it{r}} 			& $-0.281$  & $3.21\times10^{-40}$ & $-0.406 \pm 0.035$ 	& $0.674 \pm 0.059$ 	\\
{\it{i}} 			& $-0.204$  & $1.33\times10^{-21}$ & $-0.272 \pm 0.035$ 	& $0.451 \pm 0.058$ 	\\
{\it{z}} 			& $-0.147$  & $9.81\times10^{-12}$ & $-0.178 \pm 0.034$ 	& $0.296 \pm 0.058$ 	\\
{\it{J}} 			& $-0.068$  & $0.00188$ & $-0.060 \pm 0.035$ 	& $0.099 \pm 0.058$ 	\\
{\it{H}} 			& $-0.022$  & $0.322$ & $0.006 \pm 0.035$ 	& $-0.010 \pm 0.058$ 	\\
{\it{K}}$_{\text{s}}$ 	& $-0.003$  & $0.883$ & $0.032 \pm 0.035$ 	& $-0.054 \pm 0.059$ 	\\
$[3.6]$			& $-0.006$  & $0.782$ & $0.030 \pm 0.036$ 	& $-0.049 \pm 0.060$ 	\\
$[4.5]$			& $-0.001$  & $1$ & $0.039 \pm 0.036$ 	& $-0.065 \pm 0.060$ 	\\[0.5em]
\hline
\end{tabular}
\end{table*}

\begin{figure}
\includegraphics[width=0.95\columnwidth]{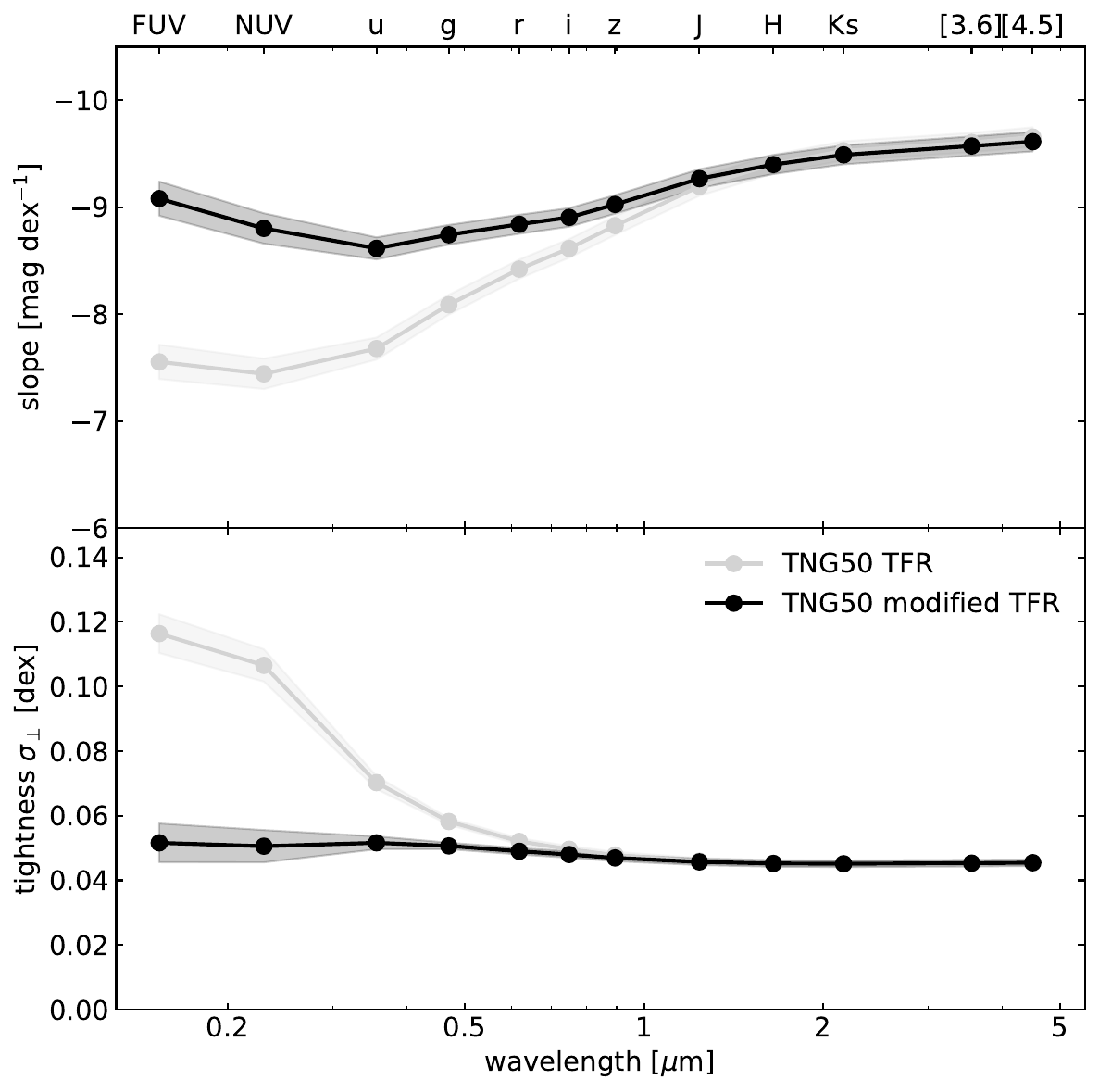}%
\caption{Slope (top panel) and tightness (bottom) panel of the modified TFR as a function of wavelength. Modified magnitudes based on the ${\textit{u}} - {\textit{r}}$ colour as a second parameter. The slope and tightness of the original TFR are shown in grey.}
\label{SlopeTightness_mod.fig}
\end{figure}

To quantify by how much a second parameter can reduce the scatter, or {{improve the tightness}}, of the TFR, we reran our fitting routine with the observed absolute magnitudes modified with the optimal linear correction. In other words, we replaced the absolute magnitudes $M$ by
\begin{equation}
M_{\text{mod}} = M + (a_Q\,\Delta M + b_Q)
\end{equation}
with $a_Q$ and $b_Q$ the slope and intercept of the best-fitting linear relation between the intensive property $Q$ and the residual from the TFR (the dotted lines in Fig.~{\ref{SecondParameter.fig}}). For the case $Q = {\textit{u}} - {\textit{r}}$, the slopes and intercepts are tabulated in Table~{\ref{ModifiedTFR.tab}} and the modified multi-wavelength TFR is shown in Fig.~{\ref{TNG50-TF-SecondParameter.fig}}. The slope, zero-point, and tightness of the multi-wavelength modified TFR is listed in Table~{\ref{TNG50-TF-mod.tab}}, and the wavelength dependence of the slope and tightness are plotted in Fig.~{\ref{SlopeTightness_mod.fig}}.

\begin{table}
\caption{Slope, zero-point, and tightness of the multi-wavelength modified TFR of our TNG50 galaxy sample. Modified magnitudes based on the ${\textit{u}} - {\textit{r}}$ colour as a second parameter.}
\label{TNG50-TF-mod.tab}
\centering
\begin{tabular}{cccc}
band & slope $a$ & zero-point $b$ & tightness $\sigma_\perp$ \\
 & [mag\,dex$^{-1}$] & [mag] & [dex] \\[0.5em]
\hline \\
FUV				& $-9.081 \pm 0.086$	& $-17.492 \pm 0.010$	& $0.052 \pm 0.002$ \\	
NUV				& $-8.802 \pm 0.081$	& $-17.710 \pm 0.009$	& $0.051 \pm 0.002$ \\	
{\it{u}}			& $-8.616 \pm 0.081$	& $-18.666 \pm 0.009$	& $0.052 \pm 0.002$ \\
{\it{g}}			& $-8.743 \pm 0.081$	& $-19.727 \pm 0.009$	& $0.051 \pm 0.002$ \\
{\it{r}}			& $-8.841 \pm 0.079$	& $-20.141 \pm 0.009$	& $0.049 \pm 0.002$ \\
{\it{i}}			& $-8.905 \pm 0.078$	& $-20.323 \pm 0.009$	& $0.048 \pm 0.002$ \\
{\it{z}}			& $-9.026 \pm 0.077$	& $-20.510 \pm 0.009$	& $0.047 \pm 0.002$ \\	
{\it{J}}			& $-9.267 \pm 0.077$	& $-20.654 \pm 0.009$ 	& $0.046 \pm 0.002$ \\	
{\it{H}}			& $-9.399 \pm 0.078$	& $-20.799 \pm 0.009$	& $0.045 \pm 0.002$ \\
{\it{K}}$_{\text{s}}$	& $-9.490 \pm 0.078$	& $-20.530 \pm 0.009$	& $0.045 \pm 0.002$ \\
$[3.6]$			& $-9.572 \pm 0.079$	& $-19.745 \pm 0.009$	& $0.045 \pm 0.002$ \\
$[4.5]$			& $-9.613 \pm 0.080$	& $-19.284 \pm 0.009$	& $0.045 \pm 0.002$ \\[0.5em]
\hline	
\end{tabular}
\end{table}

Compared to the original TFR shown in Fig.~{\ref{TNG50-TF.fig}} it can immediately be noted that the scatter in the shortest-wavelength panels is significantly reduced. This is also confirmed in the bottom panel of Fig.~{\ref{SlopeTightness_mod.fig}}. Contrary to the improving tightness with increasing wavelength of the original TFR, the tightness of the modified TFR is nearly constant at $\sigma_{\perp} \gtrsim 0.045~{\text{dex}}$ (it ranges between 0.051~dex in the FUV band to 0.046~dex in the [4.5] band). This is the value of the tightness of the original TFR in the NIR and MIR bands, for which the correction with a second parameter has not improved the tightness. The correction of absolute magnitudes by ${\textit{u}} - {\textit{r}}$ colour is responsible for nearly {\em{all}} the scatter, except for a wavelength-independent intrinsic level.

Interestingly, also the slope of the TFR changes when the absolute magnitudes are corrected (top panel of Fig.~{\ref{SlopeTightness_mod.fig}}): the slope now shows much less variation and changes only weakly from $-8.62~{\text{mag}}~{\text{dex}}^{-1}$ in the {\textit{u}} band to $-9.61~{\text{mag}}~{\text{dex}}^{-1}$ in the [4.5] band.

%%%%%%%%%%%%%%%%%%%%%%%%%%%%%%%%%%%%%%%%%%%%%%%%%%%%%%

\section{Discussion and outlook}
\label{Discussion.sec}

The main goal of this paper was to construct the multi-wavelength TFR for a set of representative disc-dominated galaxies selected from the the TNG50 simulation and check whether it can reproduce the relative behaviour of the TFR as a function of wavelength. Based on the Figs.~{\ref{SlopeTightness.fig}} and~{\ref{SlopeComparison.fig}} we can safely state that this is the case.

\subsection{Implications for the TNG50 simulation}

{{As indicated in the Introduction, reproducing the STFR or BTFR proved to be a challenge for early $\Lambda$CDM galaxy formation simulations, but recent hydrodynamical simulations with refined feedback mechanisms proved to be more successful \citep{2014MNRAS.444.1518V, 2017MNRAS.464.2419S, 2020MNRAS.498.3687G, 2021MNRAS.507.3267G, 2021A&A...651A.109D, 2023MNRAS.520.3895G}. The degree to which the STFR depends on the details of the feedback prescriptions is not completely clear. On the one hand, \citet{2014MNRAS.438.1985T} find that the STFR in the Illustris simulation \citep{2014MNRAS.444.1518V} is relatively insensitive to the feedback prescriptions; they only note that the inclusion of feedback is critical. On the other hand, \citet{2015MNRAS.450.1937C} find that the STFR is sensitive to the feedback prescriptions in the EAGLE simulation \citep{2015MNRAS.446..521S}.

\citet{2017MNRAS.464.4736F} argued that, for galaxies to both lie on the STFR and satisfy the non-monotonic galaxy--halo mass relation as determined from abundance matching \citep{2013ApJ...770...57B, 2013MNRAS.428.3121M}, their sizes need to satisfy specific constraints. This implies that the STFR is a sensitive probe of both the galaxy–halo mass relation and galaxy sizes. They show that simulated disc-dominated galaxies from the EAGLE simulation satisfy these constraints reasonably well. For the TNG50 simulation, it has also been demonstrated that the galaxies broadly agree with the abundance matching galaxy--halo mass relation \citep{2021MNRAS.500.3957E} and with the galaxy size--mass relation \citep{2019MNRAS.490.3196P, 2022MNRAS.509.2654V}, such that it is to be expected that the simulation is also consistent with the observed STFR. This on its own does not yet guarantee the good agreement with the multi-wavelength TFR that we have found here. Indeed, this also depends on the H{\sc{i}} properties, the stellar populations, and the dust attenuation characteristics, which depend on the TNG50 simulation and the radiative transfer postprocessing. The solid agreement found boosts the confidence in the TNG50 galaxy formation model and the SKIRT methodology for generating synthetic data.}}

An interesting aspect in this context is the sensitivity to resolution in the suite of TNG simulations. The TNG model was calibrated at a resolution of the original Illustris simulation \citep{2014MNRAS.444.1518V, 2014MNRAS.445..175G}, roughly equal to the resolution of the TNG100 simulation. The TNG50 simulation was run with, apart from a few minute differences, the same physical model and the same parameters for the subgrid physics recipes. Within the TNG model, an improved mass and spatial resolution results in slightly larger galaxy masses, SFRs, and luminosities \citep{2018MNRAS.473.4077P, 2018MNRAS.475..648P, 2019MNRAS.490.3196P, 2019MNRAS.485.4817D, 2022MNRAS.516.3728T, 2024MNRAS.531.3839G}. Specifically, at a given dark matter halo mass, the median stellar mass of a TNG50 galaxy is about 50\% larger than the median stellar mass of a TNG100 galaxy \citep{2021MNRAS.500.3957E}. To which degree this affects the slope and scatter of the TFR is hard to predict. One way to test this is to repeat our analysis for the TNG100 simulation, {{for which \citet{2023MNRAS.520.3895G} showed that the BTFR agrees well with the one based on the MaNGA \citep{2015ApJ...798....7B} and H{\sc{i}}-MaNGA \citep{2019MNRAS.488.3396M, 2021MNRAS.503.1345S} surveys. Such an effort would be useful to investigate the effect of resolution on the multi-wavelength TFR.}}

{{Along the same line, it would be most useful to perform a similar analysis for galaxy formation models with different physical ingredients and different feedback models. An obvious starting point could be the TNG variation volumes, which correspond to simulations with different values for many of the key parameters, including the ones related to feedback \citep{2018MNRAS.473.4077P}. Another interesting option is the recent simulation by \citet{2024arXiv240918238R}, which adds cosmic ray production and transport to the IllustrisTNG model. Cosmic rays significantly reshape the galaxy--halo mass relation while affecting galaxy sizes in a milder manner; given the importance of both of these relations in shaping the STFR \citep{2017MNRAS.464.4736F}, it would be interesting to see how the TFR relation is affected by cosmic rays. Apart from simulations from the IllustrisTNG family, our modelling approach can also be applied to other large-volume cosmological simulations such as the original Illustris, EAGLE, SIMBA \citep{2019MNRAS.486.2827D}, NewHorizon \citep{2021A&A...651A.109D}, or FIREbox \citep{2023MNRAS.522.3831F}. In principle, this could be straightforward, since the postprocessing algorithms applied in our analysis do not specifically depend on the characteristics of the hydrodynamical simulation. It falls beyond the scope of this paper, however.}}

\subsection{Towards a more stringent comparison}
\label{Towards.sec}

The fact that our TNG50 TFR reproduces the slope and tightness of the observed TFR rather well is reassuring, but the fact that different observational TFR studies predict widely varying slopes devaluates to some degree its power as a test for galaxy evolution models. In this subsection we discuss how we work towards a more stringent comparison between the observed and simulated TFR.

In its original form \citep{1977A&A....54..661T}, the TFR correlates the inclination-corrected width of the integrated H{\sc{i}} profile to the attenuation-corrected luminosity of galaxies. This is also the approach we have followed here. Other options are possible for the choice of the characteristic velocity scale, however, and it has been demonstrated that this particular choice systematically affects the slope and the scatter of the (baryonic) TFR (\citetalias{2001ApJ...563..694V}; \citealt{2007MNRAS.381.1463N}; \citetalias{2017MNRAS.469.2387P}; \citealt{2019MNRAS.484.3267L, 2023MNRAS.520.3895G}). Adopting a measure based on the line width of the integrated H{\sc{i}} velocity profile typically results in a shallower TFR than when characteristic velocities based on spatially resolved rotation curves are used. The tightest TFR is obtained when the amplitude of the rotation curve in the flat part ($v_{\text{flat}}$) is adopted as characteristic velocity. We are currently working on generating mock H{\sc{i}} data cubes for a sample of TNG50 galaxies, using an adapted version of the H{\sc{i}} post-processing of \citet{2023MNRAS.521.5645G}. From these data cubes, we are deriving rotation curves using ${}^{3{\text{D}}}$BAROLO \citep{2015MNRAS.451.3021D} in a similar way as done for real observations \citep{2024IAUGA..32P1628K}. Based on these synthetic H{\sc{i}} velocity curves, we plan to explore different measures for the characteristic velocity used in the TFR. 

Another aspect in which we can tighten the mimicking of the observational approach is the determination of the inclination. For galaxies with resolved H{\sc{i}} kinematics, the inclination can be determined from the detailed rotation curve modelling. For larger samples of galaxies without such information, the inclination is usually derived from the observed axial ratios of the isophotes in optical images \citep{1994AJ....107.2036G, 2012ApJ...749...78T, 2013AJ....146...86T}. \citet{2020ApJ...896....3K} showed that kinematically determined inclinations are slightly but systematically more face-on than the inclinations from optical images. For our galaxies, we determined the inclination directly as the angle between the direction towards the observer and the stellar angular momentum vector. We choose this option because we do not have access to the kinematic inclination. We do have synthetic optical images for about half of the sample, namely the ones with stellar masses between $10^{9.8}$ and $10^{12}~{\text{M}}_\odot$, as these galaxies are incorporated in the TNG50-SKIRT Atlas \citep{2024A&A...683A.181B}. We estimated the effect of the choice of the inclination on the TFR by replacing the angular-momentum-based inclination by the axis-ratio-based inclination for this subset of galaxies, and found essentially no difference. In future work, using the inclinations from rotation curve fits to synthetic H{\sc{i}} data cubes will further minimise differences between the observed and simulated TFR.

Finally, an important aspect that should be considered in future work is a stronger matching of the sample selection. \citetalias{2001ApJ...563..694V} demonstrated the effect of different sample selection criteria on the slope, zero-point and scatter of the TFR. These effects are substantial: for example, the slope of the {\textit{B}}-band TFR was found to vary between $-6.8\pm0.1$ for the complete H{\sc{i}} sample and $-9.0\pm0.4$ for the subsample of galaxies with a classical flat rotation curve. We also clearly illustrate the effect of the sample selection in Fig.~{\ref{SlopeTightness.fig}}: both in slope and tightness, our TNG50 TFR is sandwiched between the \citetalias{2001ApJ...563..694V} results corresponding to the SI and DE subsamples. The difference between both samples is that galaxies with bars, ongoing interactions, or disturbed morphologies are eliminated from the former sample. Turbulent pressure, non-circular motions, and out-of-equilibrium effects can cause significant deviations between the estimated rotation velocities and the actual circular velocity \citep[e.g.,][]{2020MNRAS.497.4051W, 2023MNRAS.522.3318D, 2024arXiv240416247S}. 

As discussed in Sec.~{\ref{SampleSelection.sec}}, we selected our sample to be broadly representative of the star-forming, disc-dominated galaxy population in the local Universe. However, we did not apply a further selection on the basis of H{\sc{i}} morphology or rotation curve characteristics {{(which we do not have at this moment)}}. Our sample thus most probably includes galaxies with more distorted H{\sc{i}} profiles and disturbed kinematics. In the future, with detailed H{\sc{i}} maps and ${}^{3{\text{D}}}$BAROLO rotation curves at our disposal, we will be able to select a sample with only galaxies with regular H{\sc{i}} morphologies and with high-quality and well-behaved H{\sc{i}} rotation curves. 

\subsection{Internal dust attenuation}
\label{DustAttenuation.sec}

\begin{figure*}
\centering
\includegraphics[width=0.9\textwidth]{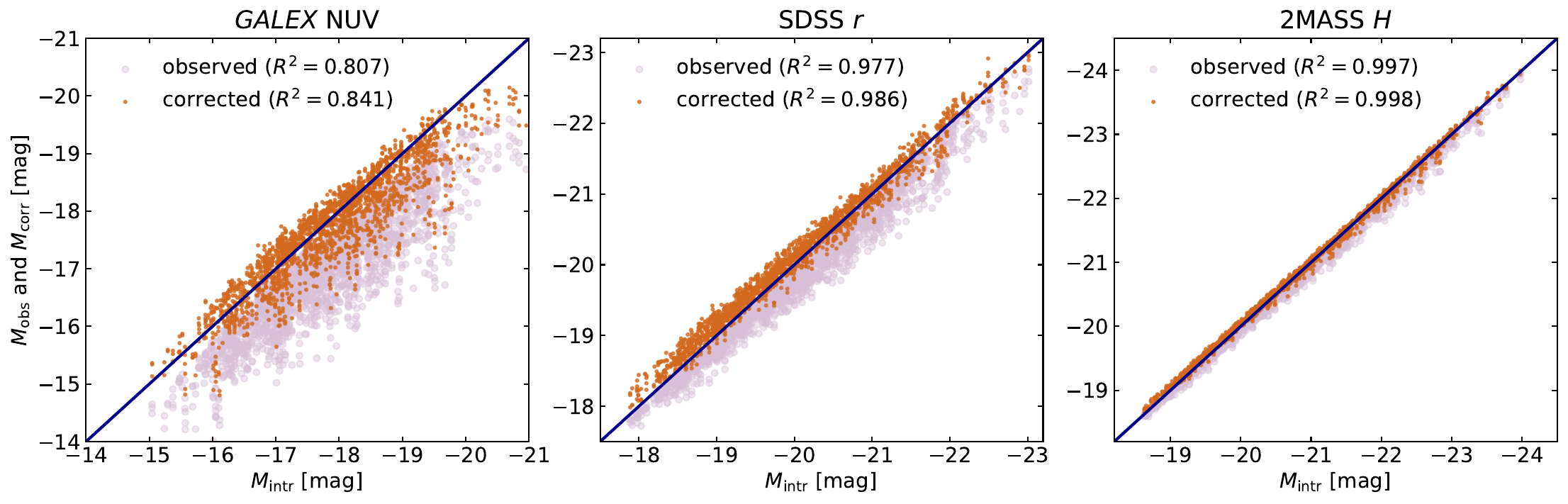}
\caption{{{Investigation of the accuracy of the attenuation correction scheme of \citet{1985ApJS...58...67T} and \citet{1996AJ....112.2471T} adopted in this paper. The three panels correspond to three different broadbands: {\it{GALEX}} NUV, SDSS {\it{r}}, and 2MASS {\it{H}}. Pink dots compare the intrinsic to the dust-attenuated absolute magnitudes for the galaxies in our TNG50 sample. Chocolate dots compare the intrinsic to the attenuation-corrected absolute magnitudes.}}}
\label{Attenuation.fig}
\end{figure*}

An important aspect in TFR studies is the correction for internal dust attenuation. As discussed in Sec.~{\ref{AbsoluteMagnitudes.sec}}, two different schemes are commonly used in TFR studies, and we decided to use the physically motivated scheme of \citet{1985ApJS...58...67T} and \citet{1996AJ....112.2471T}. As clearly illustrated by \citetalias{2001ApJ...563..694V}, the choice of the correction for internal attenuation is most uncertain and it can significantly affect the slope and scatter of the TFR relation in the optical passbands. 

Since we work with simulated galaxies, we can artificially turn off attenuation and calculate the intrinsic absolute magnitudes for each galaxy. To calculate the intrinsic absolute magnitudes, we reran SKIRT in the {\tt{NoMedium}} mode with the same settings as described in Sec.~{\ref{AbsoluteMagnitudes.sec}}. In essence, this just projects the intrinsic emission of the stellar populations on the simulated detectors without interaction with a dust medium. The result of these runs is intrinsic absolute magnitudes in the same bands and for the same orientations of all simulated galaxies in our TNG50 sample. We subsequently calculate the dust-corrected absolute magnitudes by applying the correction scheme of \citet{1985ApJS...58...67T} and \citet{1996AJ....112.2471T} to the dust-attenuated magnitudes.

The comparison of the intrinsic absolute magnitudes ($M_{\text{intr}}$), dust-attenuated or observed absolute magnitudes ($M_{\text{obs}}$), and dust-corrected absolute magnitudes ($M_{\text{corr}}$) is shown in Fig.~{\ref{Attenuation.fig}} for the {\it{GALEX}} NUV, SDSS {\it{r}}, and 2MASS {\it{H}} bands. In each panel, the pink dots compare the dust-attenuated to the intrinsic absolute magnitudes. Due to dust attenuation, the observed flux decreases, resulting in a position below the diagonal line.\footnote{In face-on view, scattering can lead to an increase of the flux, leading to an effective negative attenuation \citep{1994ApJ...432..114B, 2001MNRAS.326..733B}. Since we only consider galaxies with inclinations above 45~deg, this does not apply to the current sample}. As expected, the effect of dust attenuation is strongest in the UV and decreases with increasing wavelength \citep{2020ARA&A..58..529S}. The chocolate dots compare the dust-corrected to the intrinsic absolute magnitudes. In general, the applied correction scheme does a reasonable job in correcting for the attenuation, in the sense that the points return to a position around the diagonal line (in the ideal case, they would all exactly lie on the diagonal line). We also see a decrease of the coefficient of determination $R^2$ when we go from the intrinsic versus dust-attenuated to the intrinsic versus dust-corrected absolute magnitudes. In the {\it{r}} and {\it{H}} bands, the correction is very good for the more luminous galaxies, but the method tends to slightly over-correct the attenuation for the less luminous galaxies.

There have been recent efforts to improve on these attenuation correction schemes. Based on a sample of more than 2000 local spiral galaxies with SDSS, WISE and H{\sc{i}} data available, \citet{2019ApJ...884...82K} constructed a parametric empirical model for dust attenuation, which is subsequently used in their TFR study \citepalias{2020ApJ...896....3K}. Their empirical model depends on inclination and a single principle component that is a linear combination in roughly equal parts of H{\sc{i}} line width, a ${\text{NIR}} - {\text{H{\sc{i}}}}$ pseudo-colour, and the mean effective NIR surface brightness. 

From another angle, there are various recent and ongoing efforts to generate recipes for the attenuation based on the results of radiative transfer calculations of simulated galaxies extracted from cosmological hydrodynamical simulations \citep[e.g.,][]{2018ApJ...869...70N, 2020MNRAS.491.3937T, 2022ApJ...931...14L, 2024arXiv240111007F}. The continuous release of very realistic cosmological simulations, combined with the increased capabilities of 3D dust radiative transfer opens new avenues to improve the existing recipes. Particularly interesting in this regard are simulations that self-consistently model dust formation and evolution \citep[e.g.,][]{2018MNRAS.478.4905A, 2020MNRAS.491.3844A, 2021MNRAS.503..511G, 2022MNRAS.514.4506C, 2023ApJ...951..100N, 2024A&A...689A..79M}. We are currently working on using our SKIRT radiative transfer models based on well-resolved simulated galaxies from the TNG50 \citep{2022MNRAS.516.3728T, 2024A&A...683A.181B}, Auriga \citep{2021MNRAS.506.5703K}, and ARTEMIS \citep{2022MNRAS.512.2728C} simulations to create a new physically motivated model for the internal attenuation in spiral galaxies, and on testing the parametric attenuation model of \citet{2019ApJ...884...82K}.

%%%%%%%%%%%%%%%%%%%%%%%%%%%%%%%%%%%%%%%%%%%%%%%%%%%%%%

\section{Summary}
\label{Summary.sec}

The TFR is one of the most fundamental empirical correlations in extragalactic astronomy. Apart from its importance as a secondary distance indicator, the TFR relation serves as a test for galaxy evolution models. In this study we have derived the TFR relation in 12 broadband filters, {{ranging from {\em{GALEX}} FUV to {\em{Spitzer}} IRAC [4.5]}}, for a large sample of simulated galaxies selected from the TNG50 cosmological hydrodynamical simulation at $z=0$. For each galaxy, we have used the SKIRT radiative transfer code to generate realistic synthetic multi-wavelength global fluxes and synthetic integrated H{\sc{i}} line profiles. This enables us to derive the multi-wavelength TFR in a way similar to the approach taken by observers. Our main results can be summarised as follows.
\begin{itemize}
\item
In every broadband considered, we find a clear correlation between the absolute magnitude and the inclination-corrected H{\sc{i}} line width, which can be well approximated by a power-law relation. The TNG50 TFR systematically steepens with increasing wavelength over the NUV to MIR wavelength range. The shallowest slope ($-7.46 \pm 0.14~{\text{mag}}~{\text{dex}}^{-1}$) is obtained in the NUV band, the steepest slope ($-9.66 \pm 0.09~{\text{mag}}~{\text{dex}}^{-1}$) in the IRAC [4.5] band.
\item
{{The tightness $\sigma_\perp$ of the TNG50 TFR is worst in the FUV band ($0.116 \pm 0.022~{\text{dex}}$), and it systematically improves in the optical regime until it reaches the value of $0.046 \pm 0.001~{\text{dex}}$ in the {\em{J}} band. It remains nearly constant at this level over the entire NIR and MIR regime.}}
\item
Our TNG50 TFR reproduces the characteristic behaviour of the observed TFR in many studies: the TFR becomes steeper and tighter as we move from optical to NIR and MIR wavelengths. Quantitatively comparing our slopes to those found in observational TFR studies, we find that our results are well within the spread of different observational results. 
\item
We searched for a second parameter that can reduce the scatter in the TFR. Correlating different physical galaxy properties to the residuals of the TFR, we found no extensive property with a significant correlation strength, whereas several intensive properties were found to show moderate to strong correlations. In particular, the ${\textit{u}} - {\textit{r}}$ colour or the sSFR can significantly reduce the scatter in the UV and optical bands. Using ${\textit{u}} - {\textit{r}}$ colour as second parameter, the corrected TFR has an almost constant intrinsic tightness of about $0.046~{\text{dex}}$ over the entire UV to MIR range. 
\end{itemize}
This study is, to the best of our knowledge, the most detailed study of the multi-wavelength TFR for a large-volume cosmological hydrodynamical simulation. There are various ways in which this study can be improved or generalised, for example:
\begin{itemize}
\item
In order to compare the multi-wavelength TFR of simulated and observed galaxies in a more stringent way, the availability of H{\sc{i}} data cubes and rotation curves would be an important step forward. This will allow the construction of a better matched sample selection, the use of kinematics inclinations, and the use of different characteristic velocity scales than the integrated H{\sc{i}} profile width. A more advanced internal attenuation correction recipe would also help to limit biases.
\item
This study was based on the TNG50 simulation because it combines a large simulation volume with a relatively high mass and spatial resolution. A similar exercise for different simulations, in particular TNG100, would be useful to investigate the effect of resolution.
\item
One the fascinating prospects of new blind MeerKAT L-band surveys such as MIGHTEE-H{\sc{i}} \citep{2021A&A...646A..35M} or LADUMA \citep{2016mks..confE...4B} {{is that it is becoming possible}} to extend scaling relations to (slightly) higher redshift, and thus to explore the redshift evolution of the galaxy population in terms of their H{\sc{i}} properties \citep[e.g.,][]{2021MNRAS.508.1195P, 2022ApJ...935L..13S, 2024MNRAS.529.4192S, 2023MNRAS.525..256P}. In attendance of the SKA, which will take these relations to redshifts far beyond 1 \citep{2004NewAR..48.1013R, 2015MNRAS.450.2251Y}, extending the predictions for the TFR relations to higher $z$ is an obvious next step \citep[see also][]{2021MNRAS.507.3267G}. 
\end{itemize}

\begin{acknowledgements}
The authors thank Peter Camps and Natasha Maddox for their support of this project and the anonymous referee for helpful comments that improved the quality of the manuscript. MB, AG, SK, LL, and WJGdB gratefully acknowledge the financial support from the Flemish Fund for Scientific Research (FWO-Vlaanderen) and the South African National Research Foundation (NRF) under their Bilateral Scientific Cooperation program (grant G0G0420N). They also acknowledge the support of networking activities by NRF and the Belgian Science Policy Office (BELSPO), under grant BL/02/SA12 (GALSIMAS). MB, AG, and NA are supported by FWO-Vlaanderen through the PhD Fellowship 1193222N and Senior Research Project G0C4723N. AAP acknowledges support of the STFC consolidated grant ST/S000488/1, and the support from the Oxford Hintze Centre for Astrophysical Surveys which is funded through generous support from the Hintze Family Charitable Foundation. AS acknowledges financial support from grants CEX2021-001131-S and PID2021-123930OB-C21, both funded by MICIU/AEI/10.13039/501100011033, with additional funding from ERDF/EU. This work has received funding from the European Research Council (ERC) under the European Union's Horizon 2020 research and innovation programme (grant agreement No. 882793 ``MeerGas''). SHAR is supported by the South African Research Chairs Initiative of the Department of Science and Technology and the National Research Foundation.
This study made extensive use of the Python programming language, especially the {\tt{numpy}} \citep{Harris2020}, {\tt{matplotlib}} \citep{Hunter2007}, {\tt{scipy}} \citep{Virtanan2020}, and {\tt{astropy}} \citep{2013A&A...558A..33A} packages. 
The IllustrisTNG simulations \citep{2018MNRAS.480.5113M, 2018MNRAS.477.1206N, 2018MNRAS.475..624N, 2018MNRAS.475..648P, 2018MNRAS.475..676S} were undertaken with compute time awarded by the Gauss Centre for Supercomputing (GCS) under GCS Large-Scale Projects GCS-ILLU and GCS-DWAR on the GCS share of the supercomputer Hazel Hen at the High Performance Computing Center Stuttgart (HLRS), as well as on the machines of the Max Planck Computing and Data Facility (MPCDF) in Garching, Germany. The IllustrisTNG data used in this work are publicly available at \url{https://www.tng-project.org/}, as described by \citep{2019ComAC...6....2N}.

\end{acknowledgements}

\bibliography{mybib}

\end{document}